\documentclass[11pt]{article}
\pdfoutput=1

\usepackage[english]{babel}
\usepackage{amsmath,amssymb,amsbsy,amstext, amsthm, simplewick}
\usepackage{hyperref}
\usepackage{graphicx}
\usepackage{amsfonts}
\usepackage{amssymb}
\usepackage[small]{caption}

\makeatletter
\newlength{\apb@width}
\newcommand{\autoparbox}[2][c]{\settowidth{\apb@width}{#2}\parbox[#1]{\apb@width}{#2}}
\newcommand{\includegraphicsbox}[2][]{\autoparbox{\includegraphics[#1]{#2}}}
\makeatother

\numberwithin{equation}{section}

\def\beq{\begin{equation}}
\def\eeq{\end{equation}}

\def\bea{\begin{eqnarray}}
\def\eea{\end{eqnarray}}

\def\d{{\rm d}}

\def\beq{\begin{equation}}
\def\eeq{\end{equation}}
\def\bea{\begin{eqnarray}}
\def\eea{\end{eqnarray}}

\def\d{{\rm d}}

\DeclareRobustCommand{\SkipTocEntry}[4]{}

\begin{document}

\begin{titlepage}

\setcounter{page}{1} \baselineskip=15.5pt \thispagestyle{empty}

\bigskip\

\vspace{2cm}
\begin{center}
{\fontsize{20}{30}\selectfont  \bf Equilateral Non-Gaussianity \\ \vskip 12pt  and New Physics on the Horizon}
\end{center}

\vspace{0.5cm}
\begin{center}
{\fontsize{14}{30}\selectfont   Daniel Baumann and Daniel Green}
\end{center}


\begin{center}
\vskip 8pt
\textsl{School of Natural Sciences,
 Institute for Advanced Study,
Princeton, NJ 08540}
\end{center} 

\vspace{1.2cm}
\hrule \vspace{0.3cm}
{ \noindent \textbf{Abstract} \\[0.2cm]
\noindent 
We examine the effective theory of single-field inflation in the limit where the scalar perturbations propagate with a small speed of sound.  In this case the non-linearly realized time-translation symmetry of the Lagrangian implies large interactions, giving rise to primordial non-Gaussianities.  When the non-Gaussianities are measurable, these interactions will become strongly coupled unless new physics appears close to the Hubble scale.  Due to its proximity to the Hubble scale, the new physics is not necessarily decoupled from inflationary observables and can potentially affect the predictions of the model.  To understand the types of corrections that may arise, we construct weakly-coupled completions of the theory and study their observational signatures.}
 \vspace{0.3cm}
 \hrule

\vspace{0.6cm}
\end{titlepage}

\tableofcontents

\newpage

\section{Introduction}

Effective field theory (EFT)~\cite{Weinberg, Weinberg:1978kz} is one of the most powerful organizing principles of all 
of theoretical physics.
Its success is based on the basic observation that the physics at a particular scale of distance, time or energy doesn't depend sensitively on having detailed knowledge of the physics at  widely different scales.
The low-energy (or long-wavelength) degrees of freedom for the phenomena of interest can be isolated from the rest.
EFT makes the procedure of eliminating unnecessary high-energy degrees of freedom precise, while systematically keeping track  of their influence on the low-energy problem.
The EFT approach has been applied successfully to virtually every area of theoretical physics, but its application to cosmology is rather recent, e.g.~\cite{Cheung, Creminelli:2006xe, Senatore:2010wk}.

Given an effective theory that describes a set of experiments, one may ask when `new physics' is expected to become important.  Specifically, one would like to know what future experiments would require knowledge beyond the low-energy effective theory.  
In particle physics, future experiments typically involve colliders with sufficiently high center of mass energies $E_{\rm cm}$ to excite the new degrees of freedom. In that case, one would like to get a sense for the energy scale at which new particles are expected to be produced.  A common procedure for identifying the scale of new physics is to determine the energy scale at which the effective theory becomes strongly coupled.  As a concrete example, let us consider the Standard Model without the Higgs boson.
The low-energy effective theory with generic $W$ and $Z$ couplings becomes strongly coupled---and $WW$ scattering violates perturbative unitarity---when $E_{\rm cm} > M_W/\sqrt{\alpha} \sim 1$~TeV~\cite{Lee:1977eg}.  We therefore expect the Higgsless effective theory to break down and some form of new physics to become important at (or below) that energy scale.  Introducing a light Higgs particle, of course, eliminates the strong coupling at the TeV scale and yields an effective description that is, in principle, valid up to the Planck scale.

In this paper, we will explore an analogous situation in the context of inflationary cosmology~\cite{inflation}. 
Specifically, we will identify a regime in the effective theory of inflation~\cite{Cheung} for which the leading interactions become strongly coupled not far above the Hubble scale $H$. As in the case of the Standard Model Higgs, this suggests that new physics becomes relevant at experimentally accessible energies.

\begin{figure}[h!]
   \centering
       \includegraphics[height=6.5cm]{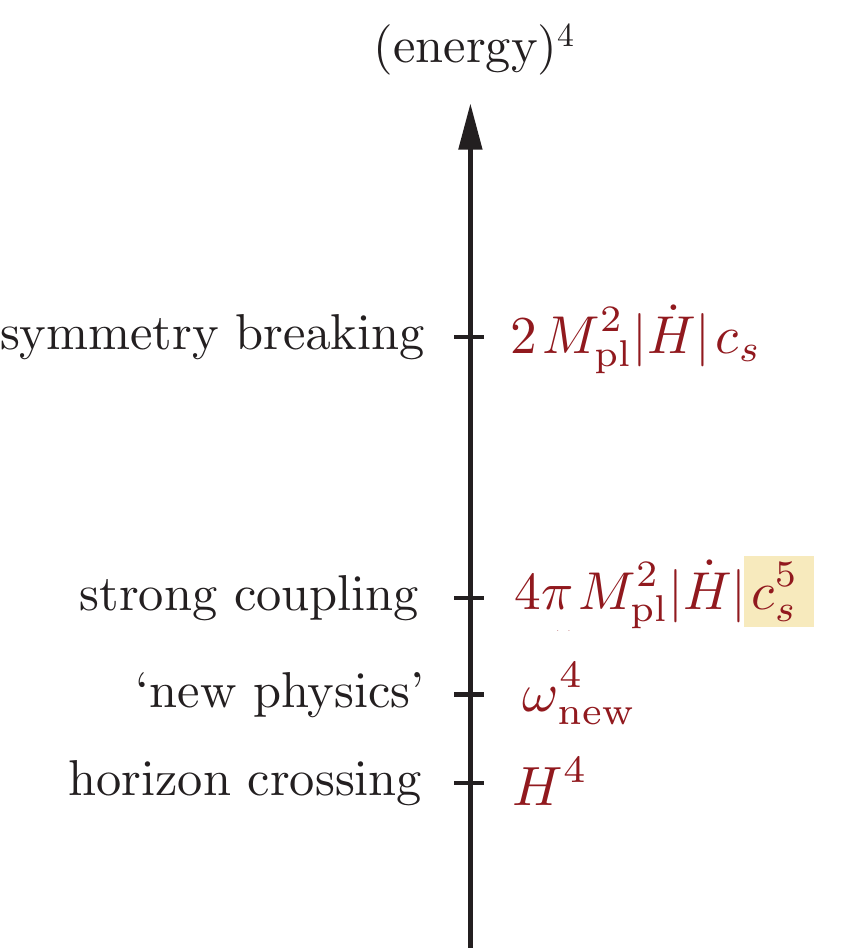}
   \caption{\sl Relevant energy scales in single-field inflation with small speed of sound.}
   \label{scales}
\end{figure}

Observations of the primordial density fluctuations, via their imprints in the cosmic microwave background (CMB), provide information about quantum-mechanical fluctuations of all fields that are lighter than the Hubble scale during inflation.  
Recently, Cheung et al.~\cite{Cheung} (see also \cite{Creminelli:2006xe}) and Senatore and Zaldarriaga~\cite{Senatore:2010wk} developed effective theories which characterize these fluctuations and their interactions at horizon crossing, i.e.~at energies near the Hubble scale.  One of these effective theories is likely to provide a complete description of future experimental data.
The theories are, in principle, valid descriptions not just at the Hubble scale but also at higher energies.  In the following, we aim to understand how and when new physics is required to alter the effective theory of single-field inflation~\cite{Cheung} when we extrapolate it to higher energies (extending our results to the effective theory of multi-field inflation~\cite{Senatore:2010wk} would be straightforward).
Just like in the example of the Higgs, we will use strong coupling as a guide to new physics. 
Since large interactions in the effective theory give rise to large non-Gaussianties of the fluctuations~\cite{Komatsu:2009kd}, strong coupling bounds are most interesting for inflationary scenarios with measurable deviations from Gaussianity.
We will find that observable equilateral non-Gaussianity implies a scale for the new physics that is not far above the Hubble scale (see Figure~\ref{scales}). 
One may then hope that the new physics is not completely decoupled and can lead to subtle signatures in the data.\footnote{At this point, it is worth remarking that the analogy between CMB and particle physics experiments is not perfect. For instance, one may rightly be concerned that future CMB experiments will not probe energy scales above Hubble directly.  In that sense, it might seem that no cosmological experiment is sensitive to the new physics. While, in principle, this is a legitimate worry, in practice, the proximity of the scale of new physics to the Hubble scale allows us to learn about the new physics without producing the new degrees of freedom. This is similar to electroweak precision tests~\cite{Peskin:1991sw}, which constrain Higgs physics without actually producing the Higgs particle.}

The effective theory of inflation \cite{Cheung} is based on the crucial insight that the inflationary perturbations are Goldstone bosons of spontaneously broken time-translation invariance of the quasi-de Sitter background.
The curvature perturbations associated with these Goldstone modes lead to the temperature anisotropies observed in the CMB. 
Since Lorentz symmetry is broken by the time-dependent background, fluctuations may propagate with a velocity (`speed of sound') that is smaller than the speed of light, $c_s \le 1$. Moreover, being Goldstone modes, the action for the perturbations is highly constrained by symmetry~\cite{Weinberg}. 
In particular, non-linearly realized time-translation symmetry relates a small value of $c_s$ to large interactions and hence large equilateral non-Gaussianities, $f_{\rm NL}^{\rm equil.} \sim c_s^{-2}$~\cite{Cheung}.
This observational signature is our prime motivation for a careful treatment of small speed of sound in the EFT of inflation.\footnote{Understanding the backgrounds that give rise to a small speed of sound in models of inflation is a well-known theoretical challenge.
Deriving a small speed of sound from a single scalar field requires that all orders in the derivative expansion of the background field $\phi$ are equally important~\cite{Creminelli:2003iq, Weinberg:2008hq}, yet be stable under radiative corrections~\cite{DBI}. This can only be achieved if the background respects a second symmetry---in addition to the shift symmetry of $\phi$ or the time-translation invariance of $H$---which protects the form of the higher-derivative interactions: In DBI inflation~\cite{DBI} the derivative expansion of the effective theory for the background is controlled by a higher-dimensional boost symmetry. In galileon inflation~\cite{Burrage:2010cu} the theory is protected by spacetime translation invariance. We will offer a few more comments about DBI inflation in Appendix~\ref{sec:DBI} 
(for related thoughts see~\cite{XingangDBI}).}

As we will show, the effective theory for single-field inflation with small $c_s$ is characterized by a special hierarchy between the fundamental energy scales of the problem (see Figure~\ref{scales}):
 the symmetry breaking scale ($\Lambda_{\rm b}$), the strong coupling scale ($\Lambda_{\star}$) and the Hubble scale ($H$).  Consistency of the effective theory requires $H$ to be the lowest of these scales.  The symmetry breaking scale is the scale at which the background is integrated out, giving a theory of the fluctuations only.  Therefore, if $\Lambda_{\star} \geq \Lambda_{\rm b}$, the effective description changes before the theory becomes strongly coupled.  
This is indeed the case for slow-roll models of inflation.
However, when $c_s \ll 1$, we find  $\Lambda_\star^4 \sim c_s^4\hskip 1pt \Lambda_{\rm b}^4 \ll \Lambda_{\rm b}^4$, so that there is a range of energies, $\Lambda_{\star} < \omega < \Lambda_{\rm b}$, where the fluctuations appear to be strongly coupled and perturbative unitarity is lost.
As in the case of particle physics, we expect new physics to become important at or below the scale where the effective theory becomes strongly coupled.  In fact, if the theory is to remain weakly coupled at all energies, the new physics will become important at energies parametrically smaller than the strong coupling scale.  The ratio $\omega_{\rm new} / \Lambda_{\star}$ reflects an expansion parameter of the weakly-coupled completion of the effective theory.

Using the measured amplitude of the power spectrum of curvature fluctuations and the relation between $c_s$ and the amplitude of non-Gaussianity $f_{\rm NL}^{\rm equil.}$, we find a strong upper bound on the scale of new physics $\omega_{\rm new}$ (see \S\ref{sec:NewPhysics}):
\beq
\label{equ:upper}
 \omega_{\rm new} \ \ll \  {\cal O}(5)\hskip 1pt \Big(\tfrac{f_{\rm NL}^{\rm equil.}}{100} \Big)^{-1/2} \, H\ .
\eeq
Here, the use of ``$\ll$" is meant to indicate that the theory is necessarily strongly coupled at the upper limit $\Lambda_\star$.  We will be interested in models that are weakly coupled even above this energy scale.  Therefore, the ratio $\omega_{\rm new} / \Lambda_{\star}$ should be sufficiently small, so that the perturbative expansion is under control.  Given that $\omega_{\rm new}$ isn't parametrically larger than the energy scale of inflation, we may hope that the effects of new physics aren't completely decoupled at $H$ and hence potentially observable.  

\vskip 4pt
The organization of the paper is as follows:
In Section~\ref{sec:Pions}, we review the effective theory of inflation~\cite{Cheung} and derive its relevant energy scales.
We argue that models with small speed of sound require new physics to appear close to the Hubble scale.
In Section~\ref{sec:NewPhysics},  in the hopes of shedding light on the nature of this new physics, we explore weakly-coupled ultraviolet (UV)-completions\footnote{We will use the term ``UV-completion" in a weaker sense than usual.  We will call a theory ``UV-complete", if the effective theory of the fluctuations is weakly coupled up to the symmetry breaking scale, at which point the background becomes important.  However, this does not mean that we have a theory for the background that would give rise to the effective theory of the fluctuations, nor have we embedded the theory in a UV-completion of gravity like string theory.} of inflationary models with small speed of sound (see also \cite{Tolley:2009fg, Cremonini:2010ua, Achucarro:2010da, Chen:2009zp}).
For each theory we determine the scale at which new physics becomes important. In Section~\ref{sec:Observ}, we compute possible observational signatures of our theories. 
As expected, for $\omega_{\rm new} > H$, the leading bispectrum is predominantly equilateral and therefore indistinguishable from the strongly-coupled `pure $c_s$--theory'. However, we identify a subleading higher-derivative  correction with a shape peaking both in the equilateral and the squashed momentum configurations.
We discuss when this signature may be detectable.
We also explain why this result motivates understanding the regime where the scale of new physics approaches the Hubble scale, $\omega_{\rm new} \to H$, and the signal in the squashed configuration becomes an order-one contribution.
In Section~\ref{sec:naturalness}, we comment on some differences concerning the notion of naturalness in our weakly-coupled UV-completions relative to the strongly-coupled effective theories.
We conclude with Section~\ref{sec:Conclusions}.

\vskip 4pt
Five appendices contain important technical details:
In Appendix~\ref{sec:DBI}, we clarify the connection between the strong coupling scale in the effective theory of the fluctuations and properties of well-known UV-completions of the background, such as DBI inflation~\cite{DBI}.  In Appendix~\ref{sec:dynamics}, we explain the dynamics of one of the UV-completions of Section~\ref{sec:NewPhysics} by examining the solutions to the equations of motion.  In Appendix~\ref{sec:Observables}, we provide details of the bispectrum calculation for our theories.  In Appendix~\ref{sec:muZero}, we present the bispectrum calculation in a novel model that arises when the scale of new physics is below the Hubble scale.  Finally, in Appendix~\ref{sec:unitarity}, we explicitly compute the strong coupling scales for all models considered in this paper.

\vskip 4pt
Throughout the text we employ the metric signature $(-,+,+,+)$ and use Greek letters $\mu, \nu, \dots$, to denote spacetime indices, while reserving Latin letters $i,j, \dots$, to label spatial indices. Spacetime indices are contracted with the metric $g_{\mu \nu}$, while spatial indices are contracted with the Kronecker delta $\delta_{ij}$.
We choose natural units with $c =\hbar = 1$ and define the reduced Planck mass as $M_{\rm pl} = (8\pi G)^{-1/2}$.
We use the letter $\pi$ to denote both $3.141\cdots$ and the Goldstone boson of broken time-translations. Which is meant should be clear from the context.

\section{The Effective Theory of Inflation}
\label{sec:Pions}

The effective theory of inflation \cite{Cheung} provides a powerful and
unified way of characterizing single-field models of inflation.
It crucially exploits the fact that inflation spontaneously breaks time-translation symmetry. 
In this section we review the effective action of the Goldstone boson associated with the spontaneous symmetry breaking.
We will explain that the theory is highly constrained by the non-linearly realized symmetries of the quasi-de Sitter background. For instance, by symmetry, a small speed of sound also implies large interactions. We will derive the strong coupling scale associated with these interactions and show that it isn't far above the Hubble scale if non-Gaussianity is to be observable.

\subsection{Goldstone Description of Inflation}

The construction of the effective theory of inflation \cite{Cheung} begins in `unitary gauge', in which there are no matter fluctuations, but only metric fluctuations.
The most general effective action is then constructed by writing down all operators that are functions of the metric fluctuations and invariant under time-dependent spatial diffeomorphisms. The two most important objects appearing in this construction are the metric perturbation $\delta g^{00}= g^{00} + 1$ and the extrinsic curvature perturbation $\delta K_{\mu \nu} = K_{\mu \nu} - a^2 H h_{\mu \nu}$, where $h_{\mu \nu}$ is the induced metric on the spatial slices. 
We use these geometrical quantities to write down the most general action with unbroken spatial diffeomeophisms~\cite{Cheung}
\bea\label{unitarygauge}
S &=& \int \d^4 x \sqrt{-g}\, \Bigl[ \, \tfrac{1}{2} M_{\rm pl}^2 R + M_{\rm pl}^2 \dot H g^{00} - M_{\rm pl}^2(3 H^2+ \dot H) \nonumber \\
&& \hspace{1.8cm} +\ \tfrac{1}{2!} M_2^4(t) (\delta g^{00})^2 + \tfrac{1}{3!} M_3^4(t) (\delta g^{00})^3 + \cdots \nonumber \\
&& \hspace{1.8cm} -\ \tfrac{1}{2} \bar M_1^3(t) \delta g^{00} \delta K^\mu_\mu - \tfrac{1}{2} \bar M_2^2(t) (\delta K^\mu_{\mu})^2 - \tfrac{1}{2} \bar M_3^2(t) \delta K^{\mu\nu} \delta K_{\mu \nu}+ \cdots \, \Bigr]\ .
\eea
Since time diffeomorphisms are broken by the time-dependence of the background, the `couplings' $H(t)$, $M_n(t)$ and $\bar M_n(t)$ are functions of time. However, the time-translation invariance is only weakly broken during inflation, so the time-dependence of these functions is typically small---e.g.~$|\dot H| \ll H^2$.
As usual in effective field theories, the action is organized as a low-energy
expansion of the fields and their derivatives: $g^{00}$ is a scalar with zero derivatives acting on it,
while $K_{\mu \nu}$ is a one-derivative object. In many situations the terms involving $g^{00}$ therefore
dominate the dynamics. 

Let us show how the effective action (\ref{unitarygauge}) unifies a large class of single-field models of inflation:\footnote{The formalism does not capture inflationary models with significant dissipative effects, such as~\cite{Berera:1995ie, Green:2009ds}. An extension of the effective theory of inflation that incorporates dissipation will appear in~\cite{Dissipation}.}
 \begin{itemize}
\item[-] The first line in (\ref{unitarygauge}) captures all single-field slow-roll models of inflation
\beq
{\cal L}_{\rm sr} = -\tfrac{1}{2} (\partial_\mu \phi)^2 - V(\phi) \ \to\  \tfrac{1}{2} \dot{\bar \phi}^2 g^{00} - V(\phi) \quad \Leftrightarrow \quad \tfrac{1}{2} \dot{\bar \phi}^2 = M_{\rm pl}^2 |\dot H| \ .
\eeq
\item[-] The second line parameterizes models with non-trivial kinetic terms 
\beq
{\cal L}_{\rm p(x)} = P(X, \phi) \ \to\ P(\dot{\bar \phi}^2 g^{00}, \bar \phi) \quad \Leftrightarrow \quad M_n^4 = \dot{\bar \phi}^{2n} \frac{\partial^n P}{\partial \bar X^n} \ .
\eeq
The operators proportional to $M_n^4$ start at order $n$ in the fluctuations. The coefficient $M_2$ induces a sound speed in the quadratic action
\beq
\label{equ:cs1}
c_s^{-2} \equiv 1 - \frac{2M_2^4}{M_{\rm pl}^2 \dot H}\ .
\eeq 
\item[-]The last line in (\ref{unitarygauge}) characterizes terms with higher derivatives that cannot be eliminated by partial integrations, such as $(\Box \phi)^2$.  Typically, these terms are suppressed by extra powers of the cutoff of the theory.  However, they can become important in cases like ghost inflation~\cite{GhostInflation} and its generalizations~\cite{LeoWMAP5}, where the leading terms vanish because $M_{\rm pl}^2 \dot H \to 0$.  
\end{itemize}

The power of the approach of~\cite{Cheung} is that it is
a completely general description for any background $H(t)$ that spontaneously breaks time-translation invariance.  
In particular, it is independent of microscopic details of the theory that gives rise to the de Sitter background.
However, as written, the dynamics of the theory are not at all clear.  
To make the dynamics more transparent, we introduce the Goldstone boson $\pi$ associated with the spontaneous breaking of time-translation invariance.
Moreover, via the St\"uckelberg trick, $\pi$ restores the full gauge-invariance of the theory.
Specifically, by definition,
the Goldstone transforms as $\pi \to \pi - \xi(x,t)$ under the time reparameterization $t \to t + \xi(x,t)$, such that $t + \pi$ is invariant.  With the replacements $t \to t + \pi$ and $g^{00} \to \partial_{\mu}(t + \pi) \partial_{\nu}(t + \pi) g^{\mu \nu}$ the action (\ref{unitarygauge}) becomes fully gauge-invariant.  For example, the slow-roll and $P(X)$ parts of (\ref{unitarygauge}) are given by  
\bea\label{invariant}
{\cal L}_{\rm sr} &=& M_{\rm pl}^2 \dot H(t+\pi) \Bigl[\partial_{\mu}(t + \pi) \partial_{\nu}(t + \pi) g^{\mu \nu} \Bigr] - M_{\rm pl}^2(3 H^2+ \dot H)(t+\pi) \ , \\
{\cal L}_{\rm p(x)} &=& \tfrac{1}{2} M_2^4(t+\pi) \Bigl[ \partial_{\mu}(t + \pi) \partial_{\nu}(t + \pi) g^{\mu \nu} + 1 \Bigr]^2 + \cdots\ . \label{equ:PX}
\eea
The quadratic term in (\ref{equ:PX}) modifies the kinetic term for the Goldstone boson, but not the gradient-squared term. It therefore leads to the sound speed cited in (\ref{equ:cs1}).
From (\ref{equ:PX}) we observe that the non-linearly realized symmetry relates a small $c_s$ (large $M_2$) to large interactions and hence observable non-Gaussianity. This limit will be of particular interest to the considerations in this paper.

Inflationary observables are often expressed in terms of the conserved curvature perturbation on comoving slices\footnote{When it was first introduced this quantity was called ${\cal R}$, in order to distinguish it from the curvature perturbation on uniform density slices, $\zeta$.  Here, we follow \cite{Maldacena:2002vr} and (mis)use $\zeta$ for the comoving curvature perturbation.}, $\zeta$. Performing a temporal gauge transformation, we relate the Goldstone boson to the comoving curvature perturbation, $\zeta = - H \pi$. Hence, up to corrections suppressed by $\frac{\dot H}{H^2}$, the correlation functions of $\pi$ are proportional to the correlation functions of $\zeta$.

\subsection{A Gauge Theory Analogy} 

The procedure of reintroducing the Goldstone boson is common in the description of massive vector bosons. Since we will make frequent use of this analogy, we digress briefly to review the gauge theory example.
Consider a non-Abelian gauge theory with Lagrangian
\beq\label{GTunitary}
{\cal L} = - \frac{1}{4}\, {\rm Tr}\, F_{\mu \nu}^2 - \frac{1}{2} m^2\, {\rm Tr}\, A_\mu^2\ .
\eeq
Under a gauge transformation with
\beq
A_\mu \to U A_\mu U^\dagger + \frac{i}{g} U \partial_\mu U^\dagger \equiv \frac{i}{g} U D_\mu U^\dagger\ ,
\eeq
the action becomes
\beq
{\cal L} =  - \frac{1}{4}\, {\rm Tr}\, F_{\mu \nu}^2 - \frac{1}{2} \frac{m^2}{g^2}\, {\rm Tr} \, D_\mu U^\dagger D^\mu U\ .
\eeq
Gauge invariance is restored by the St\"uckelberg trick, i.e.~defining $U = e^{i\pi^a T^a}$, where $T^a$ is a generator of the group.  Choosing unitary gauge, $\pi^a \equiv 0$, reproduces the action in (\ref{GTunitary}). Including $\pi$, the Lagrangian is becomes gauge-invariant and can be expanded as
\beq \label{GTpi}
{\cal L} =  - \frac{1}{4}\, {\rm Tr}\, F_{\mu \nu}^2 - \frac{1}{2} m^2 \, {\rm Tr} \, A_\mu^2 + \frac{1}{2}\frac{m^2}{g^2} (\partial_\mu \pi^a)^2 + i \frac{m^2}{g} \, {\rm Tr}\, \partial_\mu \pi^a T^a A^\mu +\ c.c.\ + \ \cdots \ .
\eeq
One of the main advantages of including the Goldstone bosons is that it makes the high-energy behavior of the theory manifest.  Specifically, it tells us that at high energies, the scattering of the longitudinal mode of the gauge field is well-described by the scattering of the Goldstone bosons.  This is most easily seen by taking the decoupling limit $m \to 0$ and $g \to 0$, while keeping $m / g \equiv f_{\pi}$ fixed.  In this limit, the Goldstone bosons decouple from $A_\mu$ and we are left with
\beq
\label{equ:NS}
{\cal L} = - \frac{1}{2} f_{\pi}^2\, {\rm Tr}\, \partial_\mu U^\dagger \partial^\mu U\ .
\eeq
Restoring finite $m$ and $g$, we should expect corrections to the results from pure Goldstone boson scattering that are perturbative in $m / E$ and $g^2$, where $E$ is the energy of the vector boson.

\subsection{Decoupling Limit}

In inflation, we would similarly like to understand the circumstances under which $\pi$ alone controls the behavior of correlation functions.  Specifically, we want to quantify the error that is made when correlation functions are computed from the {\it decoupled $\pi$-lagrangian} ${\cal L}_\pi$, i.e.~the part of the Lagrangian that doesn't include the mixing with metric fluctuations.
We can decouple the Goldstone boson $\pi$ from gravitational fluctuations by taking the limit $M_{{\rm pl}} \to \infty$ and $\dot H \to 0$, with $M_{{\rm pl}}^2 \dot H$ fixed.  This limit is equivalent to the gauge theory example via the following identifications: $ M_{\rm pl}^{-1} \leftrightarrow g$, $ \dot H \leftrightarrow m^2$ and $H \leftrightarrow E$.  Therefore, if we compute correlation functions using the decoupled $\pi$-lagrangian, our answers should be accurate up to fractional corrections of order $\frac{H^2}{M^2_{\rm pl}}$ and $ \frac{\dot H}{H^2} \equiv - \epsilon$.

We can also argue this by considering 
the dynamics of 
the comoving curvature perturbation $\zeta = - H \pi$.
During single-field inflation, the curvature perturbation freezes on superhorizon scales, $\dot \zeta \to 0$ \cite{Weinberg:2003sw}.
In other words, $\zeta$ is massless outside of the horizon.
From $\dot \zeta = - H \dot \pi - \dot H \pi$ this implies that $\pi$ cannot be precisely massless.
However, since $\pi$ is indeed massless in the decoupling limit, this mass for $\pi$ must be coming from the mixing with gravity. To compensate for the time-dependence of the Hubble rate in the relation $\zeta = -H(t) \pi$, the terms associated with the mixing with gravity such as $\delta N \dot \pi$ and $N^i \partial_i \pi$, where $N$ and $N_i$ are the standard ADM variables~\cite{ADM}, must be proportional to $\dot H$.\footnote{At least the terms that give $\pi$ a time variation outside the horizon must be proportional to $\dot H$.  In principle, mixing with gravity could also change derivative operators where the size of the effect cannot be estimated in this way.}  Note that the time-dependence of $\pi$ outside the horizon is independent of the dispersion relation.
Indeed, solving the constraint equations implied by (\ref{invariant}) and (\ref{equ:PX}), we find~\cite{Cheung:2007sv}
\beq
\delta N = \epsilon H \pi \qquad {\rm and} \qquad \partial^i N_i = - \frac{\epsilon H \dot \pi}{c_s^2}\ .
\eeq
Plugging this back into the quadratic action gives
\beq
{\cal L} = - \frac{M_{\rm pl}^2 \dot H}{c_s^2} \left( \dot \pi^2 - \frac{c_s^2}{a^2} (\partial_i \pi)^2 + 3 \epsilon H^2 \pi^2\right)\ .
\eeq
This explicitly confirms that the mass for $\pi$ arising from the mixing with gravity is $m_\pi^2 = 3 \epsilon H^2 = - 3 \dot H$.
In the remainder we will restrict to the decoupling limit,
\beq
g^{00} 
 \to -(1+\dot \pi)^2 + \frac{(\partial_i \pi)^2}{a^2} \ ,
\eeq
in which the $\pi$-lagrangian becomes
\beq
\label{eqn:decoupled}
{\cal L} \ \to\ - \frac{M_{\rm pl}^2 \dot H}{c_s^2} \left[ \dot \pi^2  - \frac{c_s^2}{a^2} (\partial_i \pi)^2 - (1-c_s^2) \Bigl( \dot \pi^3 - \dot \pi \frac{(\partial_i \pi)^2}{a^2}\Bigr) \right] + \cdots\ .
\eeq

\subsection{Energy Scales}

In order to understand the dynamics of a model, it is important to identify the energy scales at which different phenomena become important.  
Three energy scales are particularly relevant in the effective theory of inflation:
\begin{itemize}
\item[-] the {\it symmetry breaking scale}, $\Lambda_{\rm b}$, is the energy scale at which time translations are spontaneously broken and a description in terms of a Goldstone boson first becomes applicable;
\item[-] the {\it strong coupling scale}, $\Lambda_\star$, defines the energy scale at which the effective description breaks down and perturbative unitarity is lost;
\item[-] the {\it Hubble scale}, $H$, is the energy scale associated with the cosmological experiment. 
\end{itemize}
In slow-roll inflation, these three energy scales can easily be identified.  Time-translation invariance is broken by the background $\bar \phi(t)$ at the scale $\Lambda_{\rm b}^2 = \dot{\bar \phi}$.  
At energy scales above $\Lambda_{\rm b}$, the symmetry is restored and we should not integrate out the background.
Because the theory is effectively Gaussian, the self-interactions of $\phi$ are weak up to very high energies.  The theory only becomes strongly coupled at the Planck scale, so the UV-cutoff is $M_{\rm pl}$.  Inflationary observables freeze out at horizon-crossing, or when their frequencies become equal to the expansion rate, $\omega \sim H$.
Inflation therefore directly probes energies of order Hubble, i.e.~the energy scale of the experiment is $H$.  We will now define these energy scales rigorously in the effective theory, so that they can be identified in models other than slow-roll inflation.
Readers who don't want to follow the details of the derivations may jump directly to \S\ref{sec:hint} where we summarize the results and discuss their implications.

\subsubsection{Symmetry Breaking}
\label{sec:breaking}
 
Although the inflationary background spontaneously breaks a gauge symmetry, in the decoupling limit the gauge symmetry becomes a global symmetry.  As long as the decoupled $\pi$-lagrangian is a reliable description, the language of spontaneously broken global symmetries will therefore be useful.  In this section we will formulate the effective theory of inflation in a way that makes this analogy manifest.

\vskip 4pt
Let us first review the more familiar case of a spontaneously broken internal symmetry, i.e.~the decoupled Goldstone boson in a conventional gauge theory. 
By definition, any theory with a continuous global symmetry has a conserved Noether current $J^\mu$ even if  the symmetry is spontaneously broken.  There may also be an associated conserved charge
\beq
Q = \int \d^3 x \, J^{0}(x)\ .
\eeq
The existence of a well-defined $Q$ requires $J^{0}(x)$ to vanish at least as $x^{-3}$ in the limit $x \to \infty$.  In momentum space, this means that $J^{0}(p)$ scales at most like $p^{-1}$ for $p \to 0$.  
When the symmetry is spontaneously broken there is no conserved charge an infinity.  While the current always exists and satisfies $\partial_{\mu} J^{\mu} =0$, the charge itself is not well-defined.  Therefore, when the global symmetry is spontaneously broken, $J^0(x)$ will have contributions that do not fall off sufficiently rapidly as $x \to \infty$.  Nevertheless, even when the charge at infinity diverges, commutators of local fields with the charges are still well-defined.  

The current associated with the non-linear sigma model in (\ref{equ:NS}) is
\beq
\label{equ:current}
J^\mu = - f_\pi\, \partial^\mu \pi_c + \cdots\ , \qquad {\rm where} \quad f_\pi \equiv \frac{m}{g}\ .
\eeq
Here we have defined the canonically-normalized Goldstone field $\pi_c \equiv f_\pi\, \pi$.
The normalization of the current (\ref{equ:current}) is consistent with $[Q, \pi] = 1+\cdots$ (or $[Q, \pi_c] = f_\pi +\cdots$) and the commutator of the canonically-normalized field, $[\dot \pi_c({\bf x}), \pi_c({\bf y})]=i\, \delta({\bf x}- {\bf y})$.  As $x \to \infty$, we expect $\partial_0 \pi \sim x^{-2}$ by dimensional analysis and the charge therefore does not exist.  This is a direct consequence of the fact that the $\pi$ field shifts under the symmetry.

After these preliminary remarks, we will now use the current of the non-linear sigma model (\ref{equ:current})  to determine the scale at which the symmetry is broken.
We find the following two-point function\footnote{We have chosen contact terms such that $J^\mu$ is conserved when $x =0$.  This choice is necessary if one is weakly gauging the symmetry.}
\beq
\int \d^4 x\,\, e^{i p x} \, \langle 0 |  {\rm T} \{ J^{\mu}(x) J^{\nu}(0)\} | 0 \rangle =  i (p^{\mu} p^{\nu} - \eta^{\mu \nu} p^2) \, \Pi(p^2)  \ ,
\eeq
where
\beq
\label{equ:Pi}
\Pi(p^2) \equiv \frac{f^2_\pi}{p^2} + \mathcal{O}(1)\ .
\eeq
The first term in (\ref{equ:Pi}) implies $J^0 \sim p^{-2}$ and therefore the charge at infinity does not exist.  As a result, the symmetry is spontaneously broken at low energies.  The higher-order terms in (\ref{equ:Pi}) are terms that are consistent with the symmetry being unbroken.  Therefore, when $p^2 \gg f^2 $, the higher-order terms dominate and the symmetry appears to be unbroken.

\vskip 4pt
Returning to inflation, we would like to determine from (\ref{eqn:decoupled}) the scale at which time-translation invariance is spontaneously broken.  To exploit the analogy with gauge theory, we reintroduce (fake) Lorentz invariance of the action by rescaling the spatial coordinates:
\beq
\label{equ:scaled}
\tilde {\cal L}  = - \frac{1}{2} (\tilde \partial_\mu \tilde \pi_c)^2 + \cdots \ ,
\eeq
where $x \to \tilde x \equiv c_s^{-1} x$ and ${\cal L} \to \tilde {\cal L} = c_s^3 {\cal L}$.
Here we have defined the canonically-normalized field
\beq
\tilde \pi_c^2 \equiv  (2 M_{\rm pl}^2 |\dot H| c_s)\,
 \pi^2\ .
 \eeq
The Noether current associated with (\ref{equ:scaled}) is
\beq
\tilde J^\mu \equiv \tilde T^{\mu 0} = - f^2 \tilde \partial^\mu  \tilde \pi_c + \cdots\ , \qquad {\rm where} \quad f^2 \equiv (2M_{\rm pl}^2 |\dot H| c_s)^{1/2}\ .
 \eeq
The normalization of the current is consistent with $[Q, \pi] = -1$, where $Q \equiv \int \d^3 \tilde x\ \tilde T^{00}$.
Since we rescaled the spatial momenta such that everything is an energy scale, we can read off the symmetry breaking scale from our previous discussion of global currents. 
We conclude that the symmetry is spontaneously broken at
\beq
\label{equ:breaking}
 \Lambda_{\rm b}^4 \equiv 2\hskip 1pt M_{\rm pl}^2 |\dot H| c_s\ .
\eeq

One can arrive at the same result without rescaling the spatial coordinates $x$ by being careful to define an energy scale.  Recall that $T^{00} = 2\hskip 1pt M_{\rm pl}^2 \dot H c_s^{-2} \dot \pi + \cdots$ is an energy density, i.e.~an energy over volume.  Because $\dot \pi$ is dimensionless, $2\hskip 1pt M_{\rm pl}^2 \dot H c_s^{-2}$ must have units of $[\omega] [k]^3$.  To discuss energies, we use the dispersion relation, $\omega = c_s k$, to find $\Lambda_{\rm b}^4 \equiv c_s^3 (2\hskip 1pt M_{\rm pl}^2 |\dot H| c_s^{-2}) = 2\hskip 1pt M_{\rm pl}^2 |\dot H| c_s$ as before.  Using the dispersion relation to define an energy in this way will be useful in cases where the rescaling of $x$ is ineffective (cf.~\S\ref{sec:Extrinsic}).

Although the current gives a natural definition of the symmetry breaking scale, it is nice to check that it agrees with our intuition.  First of all, in the case of slow-roll inflation (i.e.~for $c_s = 1$), the symmetry breaking scale is given by $2 \hskip 1pt M_{\rm pl}^2 |\dot H| = \dot{\bar{\phi}}^2$.  This matches the intuition that the time variation of the background is breaking the symmetry.  Moreover, one may rewrite the (dimensionless) power spectrum of curvature fluctuations (cf.~\S\ref{sec:Observ}) in terms of the symmetry breaking scale
\beq
\label{equ:Pzeta}
\Delta_{\zeta} \equiv k^3 P_\zeta(k) = \frac{1}{4} \frac{H^2}{M_{\rm pl}^2 \, \epsilon\, c_s} = \frac{1}{2} \Bigl(\frac{H}{\Lambda_{\rm b}}\Bigr)^4\ .
\eeq
Hence, when $H \sim \Lambda_{\rm b}$, the size of quantum fluctuations is of the same order as the symmetry breaking scale.  This is the regime of {\it eternal inflation}, which is again consistent with the interpretation of $\Lambda_{\rm b}^4 = \dot{\bar{\phi}}^2$ for slow-roll.

\subsubsection{Strong Coupling}
\label{equ:SC}

The regime of validity of an effective theory is not always obvious.  Given a microscopic definition of the theory (i.e.~a UV-completion), the regime of validity is determined by the scales at which additional modes were integrated out.  Given only the effective description, these energy scales may not be transparent in the Lagrangian.  A fairly reliable method to identify the cutoff of the effective theory is to determine the energy scale at which the theory becomes strongly coupled. This is the approach that we will follow. Many of the results of this section are derived in detail in Appendix~\ref{sec:unitarity}.

\vskip 4pt
Let us again use the theory of massive gauge bosons as an example.  Given the action in unitary gauge (\ref{GTunitary}), it is not a priori clear where the effective description breaks down.  By carefully studying the behavior of scattering amplitudes, one finds that the scattering of the longitudinal modes of the gauge fields becomes strongly coupled at the scale $4\pi \Lambda_{\star}^2 = 4 \pi m^2 / g^2$.  This becomes more transparent after we introduce the Goldstone bosons in (\ref{GTpi}).  The action is an expansion in $\pi_c / f_{\pi}$ which contains irrelevant operators of arbitrarily large dimensions.  By dimensional analysis, the effective coupling is $\omega / f_{\pi}$ which makes the strong coupling scale at $\omega^2 = 4 \pi f^2_{\pi}$ manifest.

\vskip 4pt
Returning to the effective theory of inflation, we can similarly determine the strong coupling scale from the action for the Goldstone boson \cite{Cheung}.  For a general speed of sound, we first rescale the spatial coordinates as we did in (\ref{equ:scaled}).  The non-linear realization of the time-translation symmetry enforces relations between the quadratic, cubic and quartic actions.  Keeping only the leading interactions, we find
\beq
\tilde {\cal L} =  - \frac{1}{2} (\tilde \partial_\mu \tilde \pi_c)^2 + \frac{1}{2}\frac{(1-c_s^2)}{\Lambda_\star^2 } \dot{{\tilde \pi}}_c \frac{(\tilde \partial_i \tilde \pi_c)^2}{a^2} + \frac{1}{8} \frac{1}{\Lambda_\star^4} \frac{(\tilde \partial_i \tilde \pi_c)^4}{a^4} \ ,
\eeq
where
\beq\label{eqn:strongcoupling}
\Lambda_\star^4 \equiv 2 \hskip 1pt M_{\rm pl}^2 |\dot H| c_s^5 (1 - c_s^2)^{-1}\ .
\eeq
As in the gauge theory example, the effective coupling is given by $\omega / \Lambda_\star$.  We expect that strong coupling arises at some order-one value of this coupling.  It is useful to define the strong coupling scale by the breakdown of perturbative unitarity of the Goldstone boson scattering.   We calculate this scale in Appendix~\ref{sec:unitarity} and find that the theory is strongly coupled when $\omega^4 = 2 \pi\hskip 1pt \Lambda_\star^4$.  
Note the large suppression of $\Lambda_\star^4$ by factors of $c_s \ll 1$. Without rescaling the coordinates, the factors of $c_s$ in the strong coupling scale are less obvious.  However, the powers of $c_s$ will always agree because they convert momentum scales into energy scales.  As a result, the powers of $c_s$ are uniquely determined when we write the strong coupling scale as an energy scale. We will explain this in more detail in \S\ref{sec:NewPhysics}.

The interactions which become strongly coupled are the same that give rise to measurable non-Gaussianity.  As a result, we should be able to interpret the strong coupling scale in terms of the size of the non-Gaussianity.  A simple estimate for the amplitude of the non-Gaussianity is
\beq
\label{equ:fNLrough}
f_{\rm NL} \, \zeta\ \equiv\ \left. \frac{\mathcal{L}_3}{\mathcal{L}_2} \right|_{\omega = H} \sim \frac{M_{\rm pl}^2 \dot H c_s^{-2} (1- c_s^2)\,  \dot \pi \frac{(\partial_i \pi)^2}{a^2}}{M_{\rm pl}^2 \dot H\,  \dot \pi ^2 } = c_s^{-2}(1-c_s^2) \, \zeta  \sim \Bigl(\frac{\Lambda_{\rm b}}{\Lambda_\star}\Bigr)^2\, \zeta\ .
\eeq
Using the power spectrum (\ref{equ:Pzeta}) as an estimate for the size of $\zeta \sim \Delta_\zeta^{1/2}$, we find
\beq
\frac{{\cal L}_3}{{\cal L}_2} \sim \Bigl(\frac{H}{\Lambda_\star} \Bigr)^2\ .
\eeq
We see that $\mathcal{L}_3 \sim \mathcal{L}_2$, or $f_{\rm NL} \sim \zeta^{-1} \sim 10^{4}$, when $H \sim \Lambda_\star$. This indicates a breakdown of the perturbative description as $\Lambda_\star$ approaches $H$.  

\subsection{A Hint of New Physics?}
\label{sec:hint}

\hskip 18pt
{\it Summary.} \hskip 4pt In the previous sections we derived two important energy scales in the effective theory of inflation:   the symmetry breaking scale, $\Lambda_{\rm b}^4 = 2\hskip 1pt M_{\rm pl}^2 |\dot H| c_s$, and the strong coupling scale, $\Lambda_{\star}^4 = 2\hskip 1pt M_{\rm pl}^2 |\dot H| c_s^5 (1-c_s^2)^{-1}$.  
In slow-roll inflation, $c_s \to 1$, the strong coupling scale is much larger that the symmetry breaking scale.
However, in models with small speed of sound, $c_s \ll 1$, this hierarchy of scales is reversed, 
\beq
\frac{\Lambda_{\rm b}^4}{\Lambda_\star^4} = (1-c_s^2) c_s^{-4} \simeq 16\hskip 1pt  (f_{\rm NL}^{\rm equil.})^2  \ ,
\eeq
where we used (\ref{equ:fNLrough}) to relate $c_s$ to $f_{\rm NL}$, or more precisely $f_{\rm NL}^{\rm equil.} \approx -(4c_s^2)^{-1}$ (see \S\ref{sec:Observ}).
Therefore, any measurable non-Gaussianity ($f_{\rm NL}^{\rm equil.}\gtrsim10$) requires the strong coupling scale to appear parametrically below the scale at which the background is integrated out.

\vskip 4pt
{\it Implications.} \hskip 4pt
The inherently strongly-coupled nature of the above theories was a result of restricting the particle content of the model.  However, just like in particle physics, one should take this as an indication that new degrees of freedom may become important at energies below the scale of strong coupling.\footnote{The `new physics' could also take the form of a change in the physical description of the existing degrees of freedom. In fact, this is the case in our UV-completions in \S\ref{sec:NewPhysics}.}  Therefore, there is an energy scale $\omega_{\rm new}$ at which `new physics' becomes important.  If we wish to maintain both weak coupling and the effective small $c_s$--description at Hubble, we require $H^2 < \omega_{\rm new}^2 \ll \sqrt{2 \pi} \hskip 1pt \Lambda_\star^2$.  Given our previous results, we find
\beq
\frac{H^4}{\Lambda_\star^4} = 32 \hskip 2pt \Delta_\zeta (f_{\rm NL}^{\rm equil})^2 \ .
\eeq
where $\frac{\Delta_\zeta}{2\pi^2} = 2.4 \times 10^{-9}$ and $|f_{\rm NL}^{\rm equil.}| \lesssim 300$~\cite{Komatsu7}.
This implies that the new physics must enter not far above the Hubble scale: 
\beq
\label{eqn:bounds1}
H^2 \ < \ \omega_{\rm new}^2 \ \ll\ \sqrt{2 \pi} \hskip 1pt \Lambda_\star^2 \ \approx \  {\cal O}(20)  \hskip 1pt \Big(\tfrac{f_{\rm NL}^{\rm equil.}}{100} \Big)^{-1} \, H^2\ .
\eeq
This range of energies is sufficiently small that the new physics is not obviously decoupled at the Hubble scale.  The use of ``$\ll$" in (\ref{eqn:bounds1}) is a reminder that our loop expansion is being controlled by the ratio $\omega^2 / \sqrt{2\pi}\hskip 1pt \Lambda_\star^2$.  When $\omega_{\rm new}^2 \to \sqrt{2\pi}\hskip 1pt \Lambda_\star^2$ the theory becomes strongly coupled.  However, quantum corrections of any observable become increasingly important as one approaches this limit.  Therefore, a useful perturbative description requires that we expand in small $\omega^2 / \sqrt{2\pi}\hskip 1pt \Lambda_\star^2$, to ensure that our description is not dominated by strong dynamics.  This ratio reflects the strength of a coupling in any UV-completion of the effective theory.

\section{New Physics near Hubble}
\label{sec:NewPhysics}

In this section we will construct theories which at low energies look like small speed of sound models, ${\cal L}_{\rm sr} + \tfrac{1}{2} M_2^4 (\delta g^{00})^2$, but experience a change in their physical description at $\omega_{\rm new}^2 \ll \sqrt{2\pi} \hskip 1pt \Lambda_\star^2$, such that they remain weakly coupled up to the symmetry breaking scale $\Lambda_{\rm b}$.

\subsection{Preliminary Remarks}

The change in the physical description will not necessarily require new propagating degrees of freedom to enter at $\omega_{\rm new}$.  Instead, it may simply be the case that the scaling behavior of the field changes.  
Indeed, we will find weakly-coupled theories, in which the dispersion relation 
changes from $\omega = c_s k$ to $\omega = k^2 /  \rho$ at $\omega_{\rm new} = \rho c_s^2$, where $\rho$ is some energy scale.
The reason that this change in the dispersion relation modifies our previous result can be seen as follows:  in a relativistic theory, an operator in the action of the form $ \frac{1}{\Lambda^n} \int \d t \, \d^3 x\, \mathcal{O}_{4+n} $ implies that the strong coupling scale is $\Lambda$ if ${\cal O}_{4+n}$ is constructed from canonically-normalized fields.  However, in a non-relativistic theory, the coupling $\Lambda$ first needs to be written as an `energy scale'.  In order to do this, we have to use the dispersion relation.  For example, in the action
\beq
S= \int \d t \, \d^3 x\, \frac{a^3}{2}  \left[ \dot{\pi}^2_c - \frac{c_s^2}{a^2} (\partial_i \pi_c)^2 + \frac{1}{\Lambda^2} \, \dot \pi_c \frac{(\partial_i \pi_c)^2}{a^2}\right] \ ,
\eeq
the field $\pi_c$ has units $ [k]^{3/2} [\omega]^{-1/2}$ and therefore $\Lambda^4$ has units $[k]^{7} [\omega]^{-3}$.  
The linear dispersion relation $\omega = c_s k$ implies that the theory becomes strongly coupled at the energy scale: $\Lambda_\star^4 = \Lambda^4\, c_s^{7}$. 
For small $c_s$, the strong coupling scale $\Lambda_\star$ is therefore highly suppressed relative to the parameter $\Lambda$.
 Of course, this method for determining the strong coupling scale is equivalent to the rescaling procedure $x =c_s \tilde x$ of the previous section.

From this argument it becomes clear that by changing the dispersion relation, one can change the energy scale at which the theory becomes strongly coupled, even without changing the coefficient of the operator itself.  
Specifically, if the dispersion relation becomes non-linear before the would-be strong coupling scale is reached, $\omega \to k^2/\rho$, then the relation between the new strong coupling scale $\Lambda_\star$ and the coupling $\Lambda$ is
$\Lambda_\star^{4} = \Lambda^{4} (\Lambda/\rho)^{28}$. For $\rho^4 \ll \Lambda^4 c_s^{-1}$ this implies a larger strong  coupling scale than in the case with linear dispersion relation.
We will now present two explicit examples of weakly-coupled UV-completions that implement this change in the dispersion relation.

\subsection{Weakly-Coupled UV-Completions} 

\subsubsection{The $\pi$-$\sigma$ Model}
\label{sec:pisig}

A natural way to realize the new physics at $\omega_{\rm new}$ is by coupling $\pi$ to an additional degree of freedom\footnote{These theories are examples of the multi-field effective theory of Senatore and Zaldarriaga~\cite{Senatore:2010wk}. However, we will couple $\pi$ to fields $\sigma$ that are heavier than the Hubble scale, so we don't have to impose symmetries on the $\sigma$~fields to make them naturally light. Different types of couplings will therefore be relevant in our case. Our theories will be equivalent to the effective theory for the fluctuations in the models of~\cite{Tolley:2009fg, Cremonini:2010ua, Achucarro:2010da}.}~$\sigma$.  
 In unitary gauge we therefore consider the action
\beq
{\cal L} = {\cal L}_{\rm sr}  - \frac{1}{2}(\partial_\mu \sigma)^2 -  m^{4-n} (g^{00} + 1){\cal O}_n(\sigma) - V(\sigma)\ ,
\eeq
where ${\cal O}_n(\sigma)$ is an operator of dimension $n$ involving only the field $\sigma$. The potential $V(\sigma)$ parameterizes the self-interactions of $\sigma$. Additional couplings $(g^{00}+1)^m {\cal O}_n(\sigma)$ won't affect the quadratic action for the Goldstone $\pi$, but may be added when considering its interactions.
We could also  add derivative interactions of $\sigma$~\cite{Senatore:2010wk}.

Our goal is to generate $c_s \ll 1$ by integrating out $\sigma$. For $n > 1$, one expects the kinetic term for $\pi$ to be modified only by loops of $\sigma$.  
Therefore, let us choose ${\cal O}(\sigma) = \sigma$ and $V(\sigma) = \frac{1}{2} \mu^2 \sigma^2$.
Introducing the Goldstone boson and taking the decoupling limit, the action becomes
\beq
\label{equ:twofield}
{\cal L} = - M_{\rm pl}^2 |\dot H| (\partial_\mu \pi)^2 - \frac{1}{2} (\partial_\mu \sigma)^2 - J(\pi) \sigma - \frac{1}{2}\mu^2 \sigma\ ,
\eeq
where 
\beq
J(\pi) \equiv m^3 (g^{00}+1)\ \to\ m^3[- 2\dot \pi + (\partial_\mu \pi)^2]\ .
\eeq

\vskip 4pt
{\it Small sound speed.} \hskip 4pt 
For certain ranges of the parameters $m$ and $\mu$ (to be determined momentarily) we can integrate out $\sigma$---i.e.~$e^{i S_{\rm eff}(\pi)} = \int {\cal D}\sigma \, e^{i S[\pi, \sigma]}$---to get an effective action for the field $\pi$,
\beq
\mathcal{L}_{\rm eff} \ = \  - M_{\rm pl}^2 |\dot H| (\partial_\mu \pi)^2  + \frac{1}{2}\, J(\pi)\, (\mu^2 -\Box)^{-1} J(\pi)\ ,
\eeq
where $\Box \equiv g^{\mu \nu} \nabla_\mu \nabla_\nu \to - \partial_t^2 + \frac{1}{a^2}\partial_i^2 - 3 H \partial_t$.
Clearly, for some range of energies this is a non-local action.  
However, at sufficiently low energies a derivative expansion provides a useful description
\beq
\mathcal{L}_{\rm eff} \ = \ - M_{\rm pl}^2 |\dot H| (\partial_\mu \pi)^2  + \frac{2 m^6}{\mu^2} \dot \pi^2 + {\cal O}\Bigl(\frac{\Box}{\mu^2}\Bigr) +  \cdots  \ .
\eeq
This theory has a speed of sound given by 
\beq
\label{equ:cs2}
c_s^{-2} =1 + \frac{\rho^2}{\mu^2}\ , \qquad {\rm where} \qquad \rho^2 \equiv \frac{2m^6}{M_{\rm pl}^2 |\dot H|}\ .
\eeq
The expansion in derivatives is under control as long as $| \omega^2 - k^2 | < \mu^2$.  Since the dispersion relation is $k^2 =  \omega^2 c_s^{-2}$, the expansion is reliable when
\beq
\omega^2 < \mu^2 c_s^2 \equiv \omega_{\rm new}^2\ .
\eeq
The speed of sound can also be derived directly from the equations of motion (see Appendix~\ref{sec:dynamics}).
Although we have spoken of integrating out $\sigma$, we will now explain that {\it no} new particle appears at $\omega_{\rm new}$.\footnote{The theory is non-relativistic at low energies, so we should expect that some of the intuition derived from relativistic field theory will fail.} 
Instead, $\omega_{\rm new}$ simply marks the energy scale at which the dispersion relation changes from linear, $\omega = c_s k$, to non-linear, $\omega = k^2/\rho$.

\vskip 4pt
{\it Relation to previous works.} \hskip 4pt 
A number of previous works have recognized the possibility of using multi-field dynamics to generate a small sound speed at low energies: e.g.~in the {\it gelaton} model of
Tolley and Wyman~\cite{Tolley:2009fg} a heavy field is strongly coupled to the kinetic energy of the inflation, leading to a curved inflaton trajectory~\cite{Tolley:2009fg, Cremonini:2010ua, Achucarro:2010da}.
Integrating out the heavy gelaton field leads to an effective single-field theory with small speed of sound.
Here, we have shown how the effective theory for the fluctuations of these models arises from a systematic application of the approach of Senatore and Zaldarriaga~\cite{Senatore:2010wk}.
Next, we clarify the dynamics of these theories.

\vskip 4pt
{\it Dynamics.} \hskip 4pt 
 To discuss the dynamics of the theory at all energies between $H$ and $2 \hskip 1pt M_{\rm pl}^2 |\dot H|$, we consider the quadratic part of the two-field action 
\beq
\label{equ:two}
{\cal L}_2 = - \frac{1}{2}(\partial_\mu \pi_c)^2 - \frac{1}{2}(\partial_\mu \sigma)^2 + \rho\, \dot \pi_c \sigma - \frac{1}{2}\mu^2 \sigma^2\ ,
\eeq
where we defined the canonically-normalized field
\beq
\pi_c^2 \equiv 2 M_{\rm pl}^2 |\dot H| \, \pi^2\ .
\eeq
At energies above $\rho$, the theory is simply that of two weakly-coupled, free fields.
Below $\rho$, the mixing term $\dot \pi_c \sigma$ dominates over the kinetic terms $\dot \pi_c^2$ and $\dot \sigma^2$, and hence determines the low-energy dynamics---i.e. $\dot \pi_c \sigma$ becomes the kinetic term of the theory below $\rho$.  The terms $\dot \pi_c^2$ and $\dot \sigma^2$ are irrelevant operators in this non-relativistic theory and can be ignored at energies below $\rho$.  We can therefore drop the conventional kinetic terms and study the action
\beq
\label{equ:sig2}
{\cal L} \ \approx\ \rho\, \dot \pi_c \sigma -\frac{(\partial_i \pi_c)^2}{2\hskip 1pt a^2} - \frac{(\partial_i \sigma)^2}{2\hskip 1pt a^2} - \frac{1}{2}\mu^2 \sigma^2 + \frac{1}{\xi} \left[ \dot \pi_c^2 - \frac{(\partial_i \pi_c)^2}{ a^2} \right] \sigma + \cdots\ ,
\eeq
where
\beq
\xi \equiv \frac{2 M_{\rm pl}^2 |\dot H|}{m^3} = \frac{2}{\rho} \, (2 M_{\rm pl}^2 |\dot H|)^{1/2}\ .
\eeq
We note that, at energies below $\rho$, the two fields $\pi_c$ and $\sigma$ are {\it not} independent degrees of freedom.
In particular, at low energies, $\sigma$ plays the role of the momentum conjugate to $\pi_c$,
\beq
p_\pi \equiv \frac{\partial {\cal L}}{\partial \dot \pi_c} = \dot \pi_c + \rho \sigma \approx \rho \sigma\ .
\eeq
Hence, while the two-field action (\ref{equ:two}) describes two degrees of freedom, one of the degrees of freedom is very massive, $\omega \sim \rho$, and therefore decouples from the low-energy dynamics (this is demonstrated explicitly in Appendix~\ref{sec:dynamics}).
The dynamics of the remaining light degree of freedom are governed by the Lagrangian (\ref{equ:sig2}).
For the range of energies $\rho > \omega > \omega_{\rm new} = \mu^2 / \rho$, this describes a single degree of freedom with non-linear dispersion relation $\omega = k^2 / \rho$.  
Below $\omega_{\rm new}$, the `mass term'
$\mu^2 \sigma^2$ becomes more important than the kinetic terms and the dispersion relation becomes linear, $\omega = c_s k$, with $c_s \approx \mu/\rho \ll 1$.

\begin{figure}[h!]
   \centering
       \includegraphics[height=6.5cm]{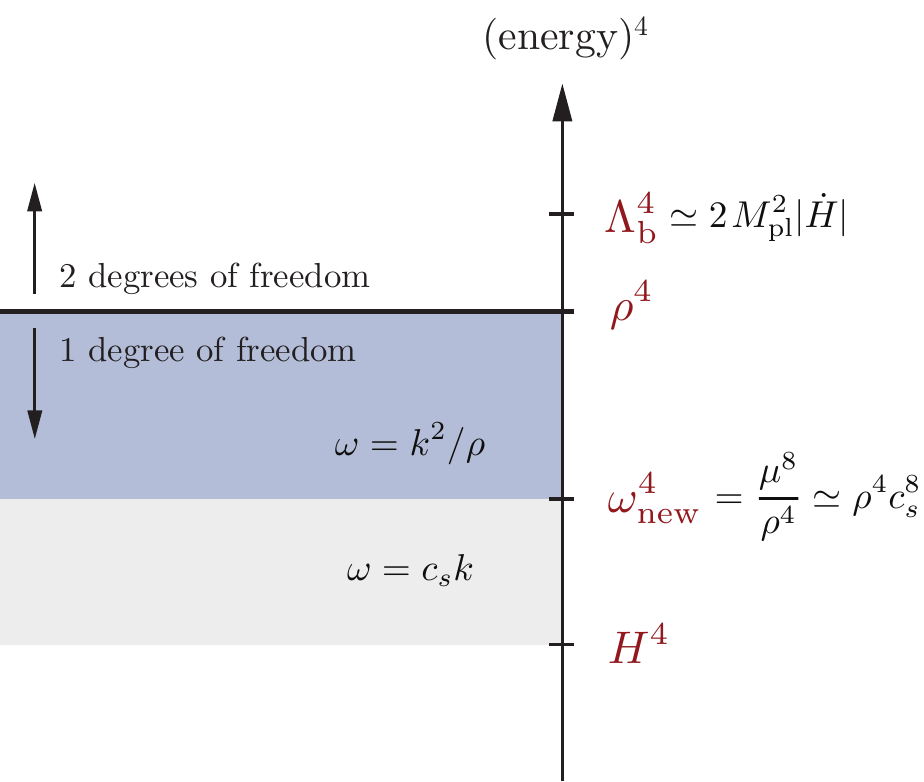}
   \label{fig:scales3}
   \caption{\sl Relevant energy scales in the $\pi$-$\sigma$ model.}
\end{figure}

\vskip 4pt
{\it Symmetry breaking.} \hskip 4pt 
Since the high-energy behavior of the theory is altered, the symmetry breaking scale will be different in the $\pi$-$\sigma$ model.
As we now show, the new symmetry breaking scale will depend on the relative size of $\rho^4$ and $2 M_{\rm pl}^2 |\dot H|$.

\vskip 2pt
For $\rho^4 < 2 M_{\rm pl}^2 |\dot H|$, the symmetry breaking scale is the same as in slow-roll inflation, 
\beq
\Lambda_{\rm b}^4 = 2 M_{\rm pl}^2 |\dot H| \qquad {\rm if} \quad \rho^4 < 2M_{\rm pl}^2 |\dot H|\ .
\eeq 
This follows simply from the fact that,
for energies above $\rho$, the theory is described by two weakly-interacting fields in a slow-roll background.  

\vskip 2pt
For $\rho^4 > 2M_{\rm pl}^2 |\dot H|$, determining the symmetry breaking scale requires more care.
Below $\rho$, we work with the effective theory (\ref{equ:sig2}).
The symmetry breaking scale in this theory is modified by the presence of the coupling $\rho \hskip 1pt \dot \pi_c \sigma$.  Specifically, the stress tensor derived from (\ref{equ:sig2}) is \ $T^{00} = \rho^{1/2} (2 M_{\rm pl}^2 |\dot H|)^{1/2}\, \sigma_c \ +\ \cdots$, where we defined the canonically-normalized field $\sigma_c \equiv \rho^{1/2} \sigma$.  Using $[ \sigma_c ] = [k]^{3/2}$ and the dispersion relation $\omega = k^2/ \rho$, we find that the symmetry breaking scale is given by $\Lambda_{\rm b}^{7/4} \rho^{3/4} = \rho^{1/2} (2 M_{\rm pl}^2 |\dot H|)^{1/2}$, or
\beq
\label{equ:sb}
\Lambda_{\rm b}^4 = 2 M_{\rm pl}^2 |\dot H| \cdot \frac{(2M_{\rm pl}^2 |\dot H|)^{1/7}}{\rho^{4/7} }  \qquad {\rm if} \quad \rho^4 > 2M_{\rm pl}^2 |\dot H|\ .
\eeq

\vskip 4pt
{\it Weak coupling.} \hskip 4pt
Finally, let us determine when our theory is weakly coupled for all energies up to the symmetry breaking scale.
To determine the cutoff associated with the Lagrangian (\ref{equ:sig2}), we define $\tilde x = \rho^{1/2} x$, $\tilde \pi_c = \rho^{-1/4} \pi_c$ and $\tilde \sigma_c = \rho^{-1/4} \sigma$,
\beq
{\cal L} \ = \ {\dot{\tilde \pi}}_c \tilde \sigma_c -\frac{(\tilde \partial_i \tilde \pi_c)^2}{2\hskip 1pt a^2} - \frac{(\tilde \partial_i \tilde \sigma_c)^2}{2\hskip 1pt a^2} + \frac{1}{2}\left[ \frac{1}{(\Lambda_\star^{(1)})^{7/4}} {\dot{\tilde \pi}}_c^2  - \frac{1}{(\Lambda_\star^{(2)})^{3/4}} \frac{({\tilde \partial_i {\tilde \pi}}_c)^2}{a^2} \right] \tilde \sigma_c\ , 
\eeq
where
\beq
(\Lambda_\star^{(1)})^{7/2} \equiv  \frac{2 M_{\rm pl}^2 |\dot H|}{\rho^{1/2}} \qquad {\rm and} \qquad  (\Lambda_\star^{(2)})^{3/2} \equiv \frac{2 M_{\rm pl}^2 |\dot H|}{\rho^{5/2}} \ .
\eeq
The scale $\Lambda_\star^{(1)}$ is always above the symmetry breaking scale (\ref{equ:sb}), so the only constraint comes from $\Lambda_\star^{(2)}$.
To avoid strong coupling before the symmetry breaking scale, we require $16\pi^2 \hskip 1pt (\Lambda_\star^{(2)})^{3/2} \gg \Lambda_{\rm b}^{3/2}$, or 
\beq
\label{equ:RHO}
\rho^4 \, \ll \, (16 \hskip 1pt \pi^2)^{7/4} \cdot 2 M_{\rm pl}^2 |\dot H|\ .
\eeq
  As we explain in Appendix~\ref{sec:unitarity}, this bound is not a unitarity bound, but arises from demanding that the loop expansion is well-defined.  

\vskip 4pt
{\it New physics.} \hskip 4pt
Where is the scale of new physics $\omega_{\rm new}^4 = \rho^4 c_s^8$ compared to the would-be strong coupling scale $2 \pi \hskip 1pt \Lambda_\star^4 \simeq 4\hskip 1pt \pi M_{\rm pl}^2 |\dot H| c_s^5$ of the small $c_s$--effective theory? 
From the above upper limit on $\rho$, cf.~Eqn.~(\ref{equ:RHO}), we derive the following bound
\beq
\label{eqn:bounds2}
\omega_{\rm new}^2 \ \ll \ 30 \cdot c_s^{3/2} \cdot   \sqrt{2 \pi} \Lambda_\star^2 \ \approx\  {\cal O}(8)  \, \Big(\tfrac{f_{\rm NL}^{\rm equil.}}{100}\Big)^{-7/4} \, H^2\ .
\eeq
We remind the reader that the ``$\ll$" in (\ref{eqn:bounds2}) reflects our requirement of a controlled perturbative expansion at $\Lambda_{\rm b}$.  We find that $\omega_{\rm new}^2$ is suppressed relative to $\Lambda_\star^2$ by powers of $c_s \ll 1$.  This is consistent with the expectation that in weakly-coupled UV-completions the new physics typically enters parametrically below the strong coupling scale.
The proximity to the Hubble scale will be relevant in the next section when we discuss the observational consequences.

Finally, we note that, for $\rho \lesssim (2 M_{\rm pl}^2 |\dot H|)^{1/4}$, the scale of new physics can even be below the Hubble scale, $\omega_{\rm new} < H$.  In this case, the dispersion relation is non-linear at horizon crossing.  At the level of the dispersion relation, the model is then similar to ghost inflation~\cite{GhostInflation}.  However, it is not the same as ghost inflation, as can be seen from the scaling dimensions of the fields.  If we assign dimensions $[t] = 1$ and $[x] = \frac{1}{2}$, then in our $\pi$-$\sigma$ model $[\pi] =[\sigma] = \frac{3}{4}$, whereas in ghost inflation $[\pi] = \frac{1}{4}$.

\subsubsection{Extrinsic Curvature Terms}
\label{sec:Extrinsic}

We have learned that a weakly-coupled UV-completion can arise if the dispersion relation changes at an energy scale below the would-be strong coupling scale in the effective theory for $\pi$.  Our $\pi$-$\sigma$ model gave an explicit realization of this idea. In this section we will present an alternative UV-completion that is weakly coupled by virtue of the same change in dispersion.  

\vskip 4pt
It is well-known that a modified dispersion relation arises in the effective theory of single-field inflation if extrinsic curvature terms are added to the action in unitary gauge~\cite{Cheung}.
Consider, for instance, the Lagrangian
\beq
\label{equ:alt}
{\cal L} \ =\ {\cal L}_{\rm sr} + \tfrac{1}{2} M_2^4 (\delta g^{00})^2 - \tfrac{1}{2} \bar M_2^2 (\delta K^\mu_\mu)^2\ ,
\eeq
where $(\delta K^\mu_\mu)^2$ is a representative extrinsic curvature term.
Going to $\pi$--gauge and taking the decoupling limit, we find
\beq
 (\delta K^\mu_\mu)^2\ \to\ \frac{(\partial_i^2 \pi)^2}{a^4} + H \frac{(\partial_i \pi)^2}{a^2} \frac{\partial_j^2 \pi}{a^2} + 2 \dot \pi \frac{\partial_i^2}{a^2} \frac{(\partial_j \pi)^2}{a^2}\ .
\eeq
For small $c_s$---i.e.~$M_2^4 \gg M_{\rm pl}^2 |\dot H|$---the quadratic action for the Goldstone mode becomes
\beq
{\cal L}_2 \ \approx\ 2 M_2^4 \left[ \dot \pi^2 - c_s^2  \frac{(\partial_i \pi)^2}{a^2} - \frac{1}{\rho^2} \frac{(\partial_i^2 \pi)^2}{a^2} \right] \ ,
\eeq
where
\beq
\rho^2 \equiv \frac{4\hskip 1pt M_2^4}{\bar M_2^2}\ .
\eeq
We see that the dispersion relation changes from linear to non-linear at $\omega_{\rm new} = \rho c_s^2$.
Hence, the strong coupling associated with the $(\delta g^{00})^2$ operator may be avoided if $\omega_{\rm new}^4 < 2 \pi \hskip 1pt \Lambda_\star^4 = 4 \pi \hskip 1pt M_{\rm pl}^2 |\dot H| c_s^5$, or 
$\rho^4 < 4 \pi M_{\rm pl}^2 |\dot H| c_s^{-3}$.

To conclude that we have a weakly-coupled model at all energies, we must check that any additional strong coupling scales induced by the new interactions in (\ref{equ:alt}) parametrically exceed the symmetry breaking scale.  First, we note that the change in the dispersion relation implies a new symmetry breaking scale $\bar \Lambda_{\rm b}$. Repeating the analysis of \S\ref{sec:breaking}, we find 
\beq
\bar \Lambda_{\rm b} = (4 M_2^4)^{2/5} \rho^{-3/5} \ . 
\eeq
 \begin{figure}[h!]
   \centering
       \includegraphics[height=6.5cm]{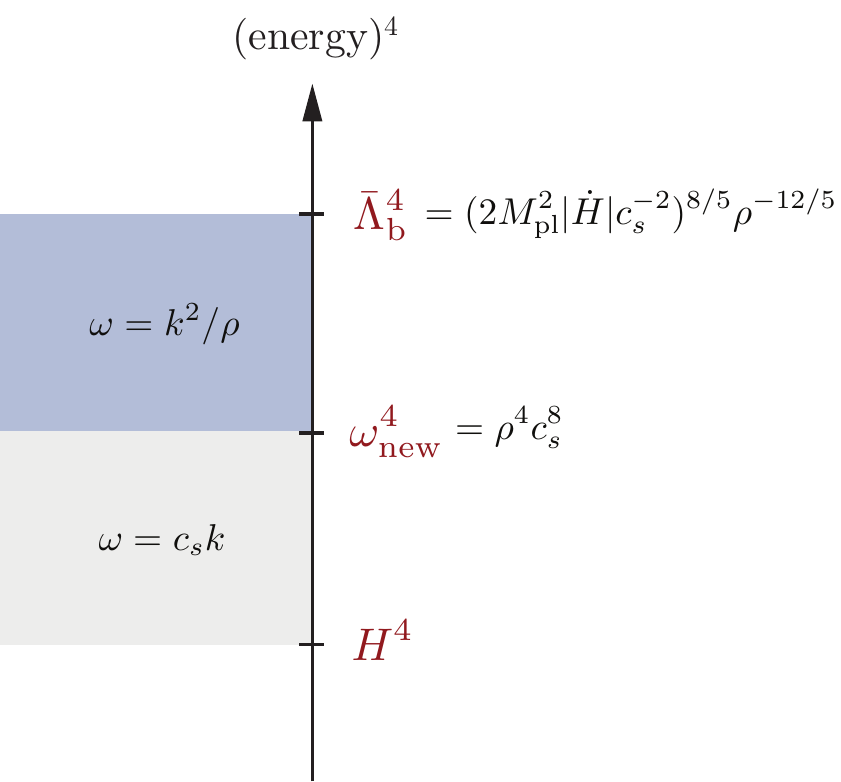}
   \label{fig:scales4}
   \caption{\sl Relevant energy scales in the model with extrinsic curvature terms.}
\end{figure}
Next, we apply the treatment of \S\ref{equ:SC} to the leading interactions in (\ref{equ:alt}).
The strong coupling scales associated with the operators $\bar M^2_2 \dot \pi \partial_i^2 (\partial_j \pi)^2$ and $M_2^4 \dot \pi^3$ are both larger than $\bar \Lambda_{\rm b}^4$ by a numerical factor.  The most stringent bound therefore comes from the operator $M_2^4 \dot \pi (\partial_i \pi)^2$, which implies that strong coupling is associated with the following energy scale
\beq
 \bar \Lambda_{\star} = (2 M_2^4)^2 \rho^{-7}\ . 
\eeq
In Appendix \ref{sec:unitarity}, we determine the order-one coefficient of the strong coupling scale from the breakdown of perturbative unitarity: $(4 \pi)^2 \hskip 1pt \bar \Lambda_{\star}$.  Requiring that $(4 \pi)^2\hskip 1pt \bar \Lambda_{\star} \geq \bar \Lambda_{\rm b}$ leads to $\rho^4 <(2 \pi)^{1/4} \hskip 1pt 4 \pi \hskip 1pt M_{\rm pl}^2 |\dot H| c_s^{-2}$, or
\beq
\omega_{\rm new} ^2 \ \ll\   \sqrt{c_s} \cdot \sqrt{2 \pi} \Lambda_{\star}^2 \ \approx\ {\cal O}(6) \, \Big(\tfrac{f_{\rm NL}^{\rm equil.}}{100}\Big)^{-5/4} \, H^2\ .
\eeq
Again, the scale of new physics is parametrically suppressed (this time only by a factor of $\sqrt{c_s}$) relative to the would-be strong coupling scale of the effective theory with only the $(\delta g^{00})^2$ operator.

\newpage
\section{Observational Consequences}
\label{sec:Observ}

We now set out to answer the question posed in the Introduction: can we detect the `new physics' that is required to enter before the strong coupling scale?
Our general strategy will be to compute the three-point functions (or bispectra) generated by our UV-completions and estimate to what extent they can be distinguished from the bispectrum of the strongly-coupled effective theory.

\subsection{{\it Pessimistic View:} Intuitive Expectations}

Although the three-point function, in principle, contains a lot of information, in practice, it is difficult to distinguish models whose dominant support is in the same momentum configuration. 
The signal-to-noise for individual modes is simply too low for a mode-by-mode comparison with the data.  Instead, the data is fit to specific templates for the three-point function with fixed momentum dependences.  
For two different templates whose dominant support lies in the same momentum configuration, the fit with either template will be of similar significance.  Two similar templates can therefore only be distinguished if the signal is detected with very high significance using either template.   
 
 Our goal is to distinguish the strongly-coupled `pure $c_s$--theory', 
 ${\cal L}_0 = {\cal L}_{\rm sr} + \tfrac{1}{2} M_2^4 (\delta g^{00})^2 $, from its weakly-coupled UV-completion.
 It is easy to see why this is a challenging undertaking.
 At low energies, the effect of the UV-completion is to add $H^2/\omega_{\rm new}^2$--suppressed derivative corrections to the `pure $c_s$--theory',
 \beq
 {\cal L} = {\cal L}_0 + \Delta {\cal L} \ ,
 \eeq
 where  $\Delta {\cal L} = {\cal O}\big(\tfrac{H^2}{\omega_{\rm new}^2}\big)$.
 It is well-known that
 ${\cal L}_0$ produces an {\it equilateral} shape~\cite{LeoWMAP5}---i.e.~the bispectrum peaks in the limit $k_1 = k_2 = k_3$.
The challenge is to pick out the subdominant correction to this background.
This will only be possible if $\Delta {\cal L}$ produces a shape that is significantly non-equilateral and hence  distinguishable from the shape produced by ${\cal L}_0$.
Since $\Delta {\cal L}$ is produced by higher-derivative corrections one may worry that its bispectrum also peaks in the equilateral configuration and is therefore degenerate with the normalization of ${\cal L}_0$.
 
Fortunately, the result of an explicit computation is a bit more optimistic than this.
In the next section, we will find that the corrections to the bispectrum arising from generic UV-completions have significant support in the {\it squashed} configuration, $k_1 = k_2 = \frac{1}{2} k_3$, and are therefore, in principle, distinguishable from the purely equilateral bispectrum.  Of course, our calculations are valid only perturbatively in $H / \omega_{\rm new}$, so the corrections are still a subdominant effect in the controlled region.  

\subsection{{\it Optimistic View:} Bispectra from Higher-Derivative Corrections}
\label{sec:explicit}

In this section, we describe the change in the bispectrum arising from new physics at $\omega_{\rm new}$.  The corrections that arise from new physics generically appear as higher-derivative operators suppressed by the scale $\omega_{\rm new}$. 
There are only a few operators that could potentially contribute to the three-point function.   For example, the corrections to the `pure $c_s$--theory' arise from terms like $\delta g^{00} \, \partial^2 \, \delta g^{00}$ or from extrinsic curvature terms.  
For concreteness, we will study the corrections in the specific case of the $\pi$-$\sigma$ model; however, one should keep in mind that these corrections will be generic to most UV-completions.  Indeed, the dominant correction to the bispectrum is identical in both the $\pi$-$\sigma$ model and the extrinsic curvature model.  Many of the details of the calculation can be found in Appendix~\ref{sec:Observables}. 

\vskip 4pt
From the expression $\omega_{\rm new} = \mu c_s = \mu^2/\rho$, we see that for fixed $c_s$ the scale of new physics can be changed continuously by altering the parameter $\mu$.  Requiring that the theory is weakly coupled at all energy scales implies an upper bound on $\mu$, but there is no lower bound (although small values may require fine-tuning).  Unfortunately, an analytic treatment isn't possible for arbitrary values of $\mu$.  However, the limits $\mu c_s \gg H$ and $\mu =0$ are both calculable and we can try to infer the general behavior from these limits.

\vskip 4pt
When $\mu c_s \gg H$, we integrate out $\sigma$ to get the following effective action for $\pi$,
\beq
{\cal L}_{\rm eff} = M_{\rm pl}^2 |\dot H| \left[ - (\partial_\mu \pi)^2 + \frac{\rho^2}{\mu^2}  \left( \dot \pi - \tfrac{1}{2} (\partial_\mu \pi)^2 \right) \left( 1 + \frac{\Box}{\mu^2} + \cdots \right) \left( \dot \pi - \tfrac{1}{2} (\partial_\mu \pi)^2 \right) \right] + \cdots\ .
\eeq
Dropping $\frac{\Box}{\mu^2}+\cdots$ reproduces the conventional small speed of sound model ${\cal L}_0$.
Since $\Box \sim H^2 c_s^{-2}$ at horizon crossing, all higher-derivative terms may be treated perturbatively as an expansion in $\frac{H^2}{\omega_{\rm new}^2}$.  
We will compute the power spectrum and the bispectrum of the comoving curvature perturbation $\zeta = - H \pi$,
\bea
\langle \zeta_{{\bf k}_1}  \zeta_{{\bf k}_2}  \rangle &=& (2\pi)^3\, P_\zeta(k_1) \, \delta({{\bf k}_1} + {{\bf k}_2})\ , \\
\langle \zeta_{{\bf k}_1}  \zeta_{{\bf k}_2}  \zeta_{{\bf k}_3} \rangle &=& (2\pi)^3 \, B_\zeta(k_1,k_2,k_3) \, \delta({{\bf k}_1} + {{\bf k}_2} + {{\bf k}_3})\ .
\eea
We reproduce the leading power spectrum for the `pure $c_s$--theory',
\beq
\Delta_\zeta \equiv k^3 P_\zeta = \frac{1}{4} \frac{H^2}{M_{\rm pl}^2 c_s \epsilon} \ .
\eeq
By rotational and translational invariance, the bispectrum is only a function of the three magnitudes $k_1$, $k_2$ and $k_3$.
Moreover, for scale-invariant fluctuations we can extract an overall factor of say $k_3^{-6}$ and write the remaining bispectrum as a function of the rescaled variables
\beq
x_1 \equiv \frac{k_1}{k_3} \qquad {\rm and} \qquad x_2 \equiv \frac{k_2}{k_3}\ .
\eeq
To compute the bispectrum up to order $\frac{H^2}{\omega_{\rm new}^2}$, there are three interactions that play a significant role,
\bea
L_{\rm int}^{(0)} &=&  {\cal C}\cdot \frac{aH}{c_s^2} \cdot\zeta' (\partial_i \zeta)^2\ , \\
L_{\rm int}^{(1)} &=&  {\cal C} \cdot \frac{H^2}{\omega_{\rm new}^2}  \cdot \zeta' \partial_i^2 \zeta'\ , \\
L_{\rm int}^{(2)} &=&  {\cal C}  \cdot \frac{H^2}{\omega_{\rm new}^2} \cdot \frac{aH}{c_s^2} \cdot\zeta' \,\frac{c_s^2 \partial_j^2}{(aH)^2}\, (\partial_i \zeta)^2\ .
\eea
where $L_{\rm int} \equiv a^4 {\cal L}_{\rm int}$ and ${\cal C} \equiv \frac{M_{\rm pl}^2 |\dot H|}{H^4} = (4\, \Delta_\zeta\, c_s)^{-1}\approx const$.
In Appendix~\ref{sec:Observables}, we compute the bispectra associated with each of these interactions.
To make contact with CMB observations, it proves useful to define an amplitude
\beq
\label{equ:fNL}
f_{\rm NL} \equiv \frac{5}{18} \frac{ B_\zeta(1,1,1)}{\Delta_\zeta^2} \ ,
\eeq
and a shape function~\cite{Babich}
\beq
S(x_1,x_2) \equiv (x_1 x_2)^2 \cdot \frac{B_\zeta(x_1, x_2,1)}{B_\zeta(1,1,1)}\ .
\eeq
To define the correlation between two distinct shapes $S$ and $S'$ we introduce the scalar product~\cite{Babich}
\beq
F(S,S') \equiv \int_{{\cal V}} S(x_1, x_2) S'(x_1, x_2) \, \omega(x_1, x_2) \, \d x_1 \d x_2\ ,
\eeq
where the integrals are only over physical momenta satisfying the triangle inequality: $0 \le x_1 \le 1$ and $1-x_1 \le x_2 \le 1$.
We introduced a weight function in the integral,
 $\omega(x_1, x_2) \equiv (1+x_1+x_2)^{-1}$, to achieve that the scalar product exhibits the same scaling as the optimal CMB estimator~\cite{Shellard}.
As a measure of the degree of correlation between two shapes we use the normalized scalar product or `cosine'~\cite{Babich}
\beq
\label{equ:cosine}
{\cal C}(S, S') \equiv \frac{F(S,S')}{\sqrt{F(S,S) F(S',S')}}\ .
\eeq

The leading non-Gaussianity is generated by $L_{\rm int}^{(0)}$, the interaction associated with the operator $(\delta g^{00})^2$.
Computing the shape of the bispectrum in the in-in formalism (see Appendix~\ref{sec:Observables}) gives~\cite{LeoWMAP5}
\beq
S^{(0)}\ =\ - \frac{9}{17} \cdot \frac{ X_1^{\, 6} - 3 X_1^{\, 4} X_2^{\, 2}  + 11 X_1^{\, 3}  X_3^{\, 3}    - 4 X_1^{\, 2} X_2^{\, 4}  - 4 X_1  X_2^{\, 2} X_3^{\, 3}     + 12 X_3^{\, 6}   }{X_3^{\, 3} X_1^{\, 3}}\ ,
\eeq
where 
\bea
X_1 &=& 1+ x_1 +x_2\ , \\
X_2 &=& (x_1 x_2 +x_1 + x_2)^{1/2} \ ,\\
X_3 &=& (x_1 x_2)^{1/3}\ .
\eea
The shape $S^{(0)}$ peaks in the equilateral configuration $x_1=x_2=1$. In fact, we will use $S^{(0)}$ as our definition of the `equilateral shape' $S_{\rm equil} \equiv S^{(0)}$.  
The amplitude of this equilateral non-Gaussianity is~\cite{LeoWMAP5}
\beq
f_{\rm NL}^{\rm equil.} \equiv f_{\rm NL}^{(0)} = \frac{85}{324} \left(1 - \frac{1}{c_s^2} \right) \simeq - \frac{1}{4c_s^2}\ .
\eeq
A second shape of interest arises as a special linear combination of the shapes associated with the operator $M_2^4(\delta g^{00})^2$ and the operator $M_3^4 (\delta g^{00})^3 \to  \tilde c_3 \cdot {\cal C}\cdot \frac{a H}{c_s^4} (\zeta')^3$.
Both operators individually produce similar (but not identical) equilateral shapes.
By tuning the coefficient $\tilde c_3$, we define an `orthogonal shape'~\cite{LeoWMAP5} by the value of $\tilde c_3$ for which the correlation between the equilateral shape $S^{(0)} \equiv S_{\rm equil}$ and the orthogonal shape $S_{\rm ortho}$ vanishes
\beq
{\cal C}(S_{\rm equil}, S_{\rm ortho}) \equiv 0\ .
\eeq
This occurs when $\tilde c_3 \simeq -5.4$.

Both $L_{\rm int}^{(1)}$ and $L_{\rm int}^{(2)}$ contribute to the bispectrum at leading order in $\frac{H^2}{\omega_{\rm new}^2}$.  The interaction $L_{\rm int}^{(1)}$ furthermore leads to a small correction to the power spectrum  (see Appendix~\ref{sec:Observables}). However, this correction is scale-invariant and so only corresponds to an unobservable shift in the amplitude.
One may hope that the shape of the correction to the bispectrum, $S^{(1)}$, leaves a more detectable imprint.
However, we find that $S^{(1)}$ is not significantly different from the equilateral shape $S^{(0)}$. In fact, the cosine between the shape $S^{(1)}$ and the orthogonal shape is only
\beq
{\cal C}(S^{(1)}, S_{\rm ortho}) = 0.21 \ .
\eeq
As a result, the correction to the bispectrum from $L_{\rm int}^{(1)}$ will, in practice, be difficult to distinguish from a shift in the value of $M_2$.

\begin{figure}[h!]
   \centering
       \includegraphics[width=.85\textwidth]{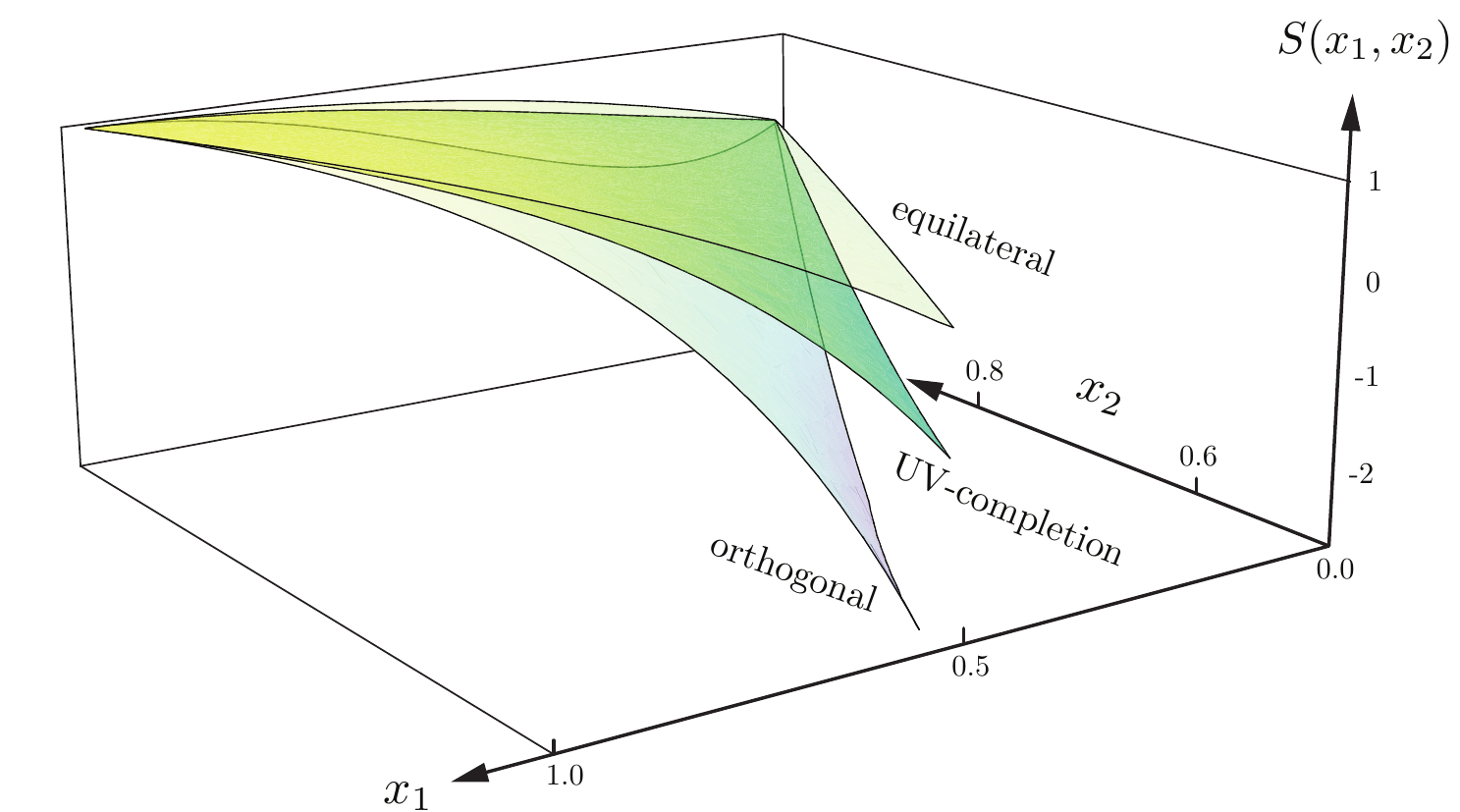}
   \caption{\sl Comparison between the shape produced by the dominant higher-derivative correction in both UV-completions, the equilateral shape and the orthogonal shape. All shapes are normalized relative to the amplitude in the equilateral configuration.}
      \label{fig:S2}
\end{figure}

The more interesting correction comes from $L_{\rm int}^{(2)}$.  This is the same interaction as appears in the extrinsic curvature term described in \S\ref{sec:Extrinsic}.    
The shape of the corresponding contribution to the bispectrum is
\bea
\label{eqn:shapecorrection}
&& S^{(2)} \ =\  \frac{27}{13} \cdot \frac{X_1^{\, 8}  - 3 X_1^{\, 6} X_2^{\, 2}  - 7 X_1^{\, 4} X_2^{\, 4} +12 X_1^{\, 2} X_2^{\, 6} + 17 X_1^{\, 5} X_3^{\, 3}   }{X_3^{\, 3} X_1^{\, 5}} \, \times \nonumber \\
&& \hspace{1.8cm}\times \
 \frac{ - 43 X_1^{\, 3} X_2^{\, 2} X_3^{\, 3} + 36 X_1 X_2^{\, 4} X_3^{\, 3} + 66 X_1^{\, 2} X_{3}^{\, 6} - 48 X_2^{\, 2} X_3^{\, 6} }{X_3^{\, 3} X_1^{\, 5}}\ .
\eea
The shape $S^{(2)}$ shows a significant peak in the squashed configuration (see Figure~\ref{fig:S2}).\footnote{The same shape also appears in the effective theory of single-field inflation if certain extrinsic curvature operators are considered~\cite{Bartolo:2010bj}.}
This is reflected in the cosine between this shape and the orthogonal shape 
\beq
{\cal C}(S^{(2)}, S_{\rm ortho}) = 0.56\ .
\eeq
This overlap with the orthogonal shape gives us hope that $S^{(1)}$ could be detectable in a measurement using the orthogonal template.
However, although quite distinct in shape, this correction will be very hard to extract from the data if the amplitude of the correction is small. This is required by our wish to maintain perturbative control over our calculations, but does not have to be the case more generally.

\subsection{{\it Realistic View:} Observational Prospects}

In the previous section, we computed the correction to the bispectrum produced by physics near the Hubble scale.  
We found that the correction $S^{(2)}$ has a significant overlap with the orthogonal shape, which by definition has zero overlap with the equilateral shape $S^{(0)}$. 
When we measure non-Gaussianity using the orthogonal template, we are therefore effectively projecting out the contribution to the signal from $S^{(0)}$.
In the `pure $c_s$--theory', the signal is only coming from $S^{(0)}$, so we don't expect to measure $f_{\rm NL}^{\rm ortho.}$ in that case.
In contrast, in our UV-completions, we can get contributions both to $f_{\rm NL}^{\rm equil.}$ (mostly from $S^{(0)}$) and to $f_{\rm NL}^{\rm ortho.}$ (mostly from $S^{(2)}$).
We propose to use this correlated signature as a diagnostic for our UV-completions.
In this section, we will use this fact to give a rough estimate for when the contribution $S^{(2)}$ can be measured in future experiments.

\vskip 4pt
The shape computed in (\ref{eqn:shapecorrection}) is the first term in a perturbative expansion in $\frac{H^2}{\omega_{\rm new}^2}$.  Therefore, the amplitude of this contribution is suppressed by $\frac{H^2}{\omega_{\rm new}^2} < 1$.  
From the calculation in Appendix~\ref{sec:Observables}, we infer that the relative size of the two bispectra in the equilateral limit is 
\beq
B^{(2)}_\zeta(1,1,1) =  \frac{26}{51} \cdot \frac{H^2}{\omega_{\rm new}^2} \cdot B_\zeta^{(0)}(1,1,1)\ .
\eeq
Using the standard definition of $f_{\rm NL}$ (\ref{equ:fNL}),
we find that the relative contribution to $f_{\rm NL}$ coming from the higher-derivative correction is given by
\beq
\label{eqn:correctionfnl1}
f_{\rm NL}^{(2)} \simeq  \frac{1}{2}\frac{H^2}{\omega_{\rm new}^2} f_{\rm NL}^{(0)} \ .
\eeq
In practice, we do not measure this $f_{\rm NL}$, but rather use the equilateral and orthogonal templates to determine $f_{\rm NL}^{\rm equil.}$ and $f_{\rm NL}^{\rm ortho.}$.  The contribution from (\ref{eqn:correctionfnl1}) to $f_{\rm NL}^{\rm equil.}$ can be absorbed by a change in the normalization of $L_{\rm int}^{(0)}$ (the size of $c_s$).\footnote{In the perturbative regime we will use $f_{\rm NL}^{\rm equil.} \simeq f_{\rm NL}^{(0)}$.}
We will therefore focus on $f_{\rm NL}^{\rm ortho.}$.

The definition of the cosine in (\ref{equ:cosine}) was chosen to related the experimental bounds from different templates \cite{Babich}. This allows us to estimate the contribution from (\ref{eqn:correctionfnl1}) to $f_{\rm NL}^{\rm ortho.}$ as
\beq
\Delta f_{\rm NL}^{\rm ortho.}\ \simeq \ \frac{1}{2}\, {\cal C}(S^{(2)}, S_{\rm ortho}) \cdot \sqrt{\frac{S_{\rm ortho.}\cdot S_{\rm ortho.} }{S_{\rm equil.}\cdot S_{\rm equil.} } } \cdot \frac{H^2}{\omega_{\rm new}^2} f_{\rm NL}^{\rm equil.}  \ .
\eeq
Using $ {\cal C}(S^{(2)}, S_{\rm ortho}) = 0.56$ and $S_{\rm ortho}\cdot S_{\rm ortho} \simeq \frac{9}{4} \hskip 1pt S_{\rm equil.} \cdot S_{\rm equil.}$, we get
\beq
\label{eqn:fnlcorrection}
\Delta f_{\rm NL}^{\rm ortho.} \sim  \frac{1}{2} \frac{H^2}{\omega_{\rm new}^2} f_{\rm NL}^{\rm equil.} \ .
\eeq
This is the size of the predicted non-Gaussianity measured with the orthogonal template if the non-Gaussianity measured using the equilateral template is $f_{\rm NL}^{\rm equil}$.

As we explained before---cf.~(\ref{eqn:bounds2})---the ratio $H^2/\omega_{\rm new}^2$ satisfies 
\beq
\label{equ:ratio}
\frac{1}{8}\, \Big(\tfrac{f_{\rm NL}^{\rm equil.}}{100}\Big)^{7/4}  \ \ll \ \frac{H^2}{\omega_{\rm new}^2}\ <  \ 1  \ .
\eeq
Combining (\ref{eqn:fnlcorrection}) and (\ref{equ:ratio}), we find 
\beq
\label{equ:signal}
6 \cdot \Big(\tfrac{f_{\rm NL}^{\rm equil.}}{100} \Big)^{11/4} \ \ll \ \Delta f_{\rm NL}^{\rm ortho.}  \ < \ 50 \cdot \Big(\tfrac{f_{\rm NL}^{\rm equil.}}{100} \Big)\ .
\eeq
Interestingly, we get a lower bound on the expected signal from the requirement that the theory be perturbative at $\Lambda_{\rm b}$---which is responsible for the ``$\ll$" in (\ref{equ:ratio}).
However, we emphasize that
we are pushing the validity of the perturbative calculation as we approach the upper limit in (\ref{equ:signal}). 
For the contribution to the orthogonal shape to be detectable in the regime of perturbative control (i.e.~$\Delta f_{\rm NL}^{\rm ortho.} \gtrsim 10$ with $H < \omega_{\rm new}$), we require a detection of equilateral non-Gaussianity near its current upper limit, $|f_{\rm NL}^{\rm equil.}| \lesssim 250$~\cite{Komatsu7}.
Even in this optimistic case, the signal will be hard to detect with future CMB experiments.\footnote{To get a sense for the significant observational challenge that this implies, we remind the reader of the WMAP 95\% C.L.~constraints~\cite{Komatsu7}: $-214 < f_{\rm NL}^{\rm equil.} < 266$ and $-410 < f_{\rm NL}^{\rm ortho.} < 6$. Forecasted 1-$\sigma$ errors for future experiments are~\cite{Liguori:2010hx, Amit}:
$\sigma( f_{\rm NL}^{\rm equil.}, f_{\rm NL}^{\rm ortho.}) \sim 30$ (Planck~\cite{Planck}) and $\sigma( f_{\rm NL}^{\rm equil.}, f_{\rm NL}^{\rm ortho.}) \sim 10$ (CMBPol~\cite{CMBPOL}, COrE~\cite{CORE}).}

\vskip 4pt
The above conclusions relied on limiting ourselves to the perturbative regime, $H \ll \omega_{\rm new}$.
Given the strong upper bound on the scale of new physics, the more likely scenario is when $\omega_{\rm new} \sim H$, where our perturbative techniques break down.  The perturbative calculation may be suggestive that this limit could generate a significant contribution to $f_{\rm NL}^{\rm ortho.}$.
This is a well-motivated scenario that would become relevant in the event of a detection of equilateral non-Gaussianity by the Planck satellite. 

\subsection{The $\omega_{\rm new} \to 0$ Limit}
\label{sec:muZeroX}

The calculation of the previous section suggests that a significant orthogonal component could be generated as $\omega_{\rm new} = \mu^2/\rho \to H$.  Unfortunately, we were unable to explicitly calculate the bispectrum in this limit. 
In the absence of an analytic bispectrum calculation for all values of $\mu$, it is still interesting to study the extreme limit, $\mu \to 0$.  This limit is also interesting simply because it is a new single-field model whose dynamics haven't previously been considered in the literature.  In this section, we will determine the observational signatures of the $\mu \to 0$ limit of the $\pi$-$\sigma$ model, with many details left to Appendix~\ref{sec:muZero}.

\vskip 4pt
The Lagrangian describing this limit is 
\beq\label{eqn:muzero}
{\cal L} \ \approx\ \rho\, \dot \pi_c \sigma - \frac{1}{2} \frac{(\partial_i \pi_c)^2}{a^2} - \frac{1}{2}\frac{(\partial_i \sigma)^2}{a^2}  \underbrace{- \frac{1}{\xi} (\partial_\mu \pi_c)^2 \sigma}_{{\cal L}_{\rm int}}\ ,
\eeq
where $\xi \equiv \frac{2 M_{\rm pl}^2 |\dot H|}{m^3} = \frac{2}{\rho}\, (2 M_{\rm pl}^2 |\dot H|)^{1/2}$. The interaction ${\cal L}_{\rm int}$ will be responsible for generating a measurable bispectrum.  Focusing first of the quadratic action, we get the following equations of motion
\beq
 \ddot \pi_c + 5 H \dot \pi_c + \frac{k^4}{\rho^2 a^4}\, \pi_c = 0 \qquad \,{\rm and} \, \qquad \frac{k^2}{a^2} \, \sigma =  \rho\, \dot{\pi}_c\ .
\eeq
Note the unusual Hubble friction factor of $5H$ (rather than $3 H$) that occurs in this model.  This is a consequence of $\sigma$ (rather than $\dot \pi_c$) being the canonical momentum of $\pi_c$.  Both of these facts are important for achieving a scale-invariant power spectrum.

The mode functions are most easily determined in conformal time. 
As usual, we will quantize the system by writing $ \hat \pi_c ({\bf k}, \tau) = \pi_k(\tau) \hat a_{\bf k} + \pi^*_k(\tau) \hat a^\dagger_{- {\bf k}}$, where $\pi_k(\tau)$ is a solution to the equations of motion.  Being careful to define the Bunch-Davies vacuum when $\sigma$ is the canonical momentum (see Appendix~\ref{sec:muZero}), one finds
\beq
\pi_k (\tau) = (H \tau)^2 \, \frac{\sqrt{- k^2 \tau}}{\rho} \, H_{5/4}^{(1)}\Bigl(\tfrac{1}{2}\tfrac{H}{\rho} (k \tau)^2 \Bigr)\ ,
\eeq
where $H^{(1)}_{\nu}(x)$ is the Hankel function of the first kind.  The resulting power spectrum for $\zeta$ is scale-invariant, with amplitude
\beq
 \Delta_\zeta \equiv k^3 |\zeta_k^{(o)}|^2 \sim \frac{H^2}{M_{\rm pl}^2 \epsilon}  \left( \frac{\rho}{H}\right)^{1/2}  \ .
 \eeq
 The bispectrum calculation proceeds as usual if we use the interaction ${\cal L}_{\rm int} = - \frac{1}{\xi} (\partial_\mu \pi_c)^2 \sigma$, with $\sigma =  - \frac{\rho}{H \tau k^2}\,  \pi_c'$.  Unfortunately, the bispectrum can only be computed numerically.  The resulting shape is very similar to the one generated by the $M_2^4\, \dot \pi (\partial_\mu \pi)^2$ interaction in the `pure $c_s$--theory',
 \beq
 {\cal C}(S_{\mu=0}, S^{(0)}) = 0.99\ .
 \eeq
The similarity of the shapes can be understood at the level of the interaction Hamiltonian. Because $\sigma \propto \dot \pi / k^2$, the basic kinematic momemtum factor is still $({\bf k}_1 \cdot {\bf k}_2 + {perms.})$, arising from the $\partial_i \pi\, \partial^i \pi$--part of ${\cal L}_{\rm int}$.  Moreover, although the mode functions are different, they are still functions of $k$ that vanish exponentially inside the horizon.  In contrast, the corrections computed in the previous section included interactions of the form $\dot \pi\, \partial_i \partial_j \pi\, \partial^i \partial^j \pi$.  The resulting kinematic factor therefore differs and introduces a significant signal in the squashed momentum configuration.  

The limit $\mu \to 0$ ($\omega_{\rm new} \to 0$) in the $\pi$-$\sigma$ model has an analogue in the extrinsic curvature model discussed in \S\ref{sec:Extrinsic}.  In that case, the limit $\omega_{\rm new} \to 0$ corresponds to the limit $\dot H \to 0$ with finite $M_2$ and $\bar{M}_2$.  This model is typically referred to as ``ghost inflation"~\cite{GhostInflation}.  The predictions of ghost inflation are discussed in detail in \cite{LeoWMAP5}.  
Its correlation with the orthogonal shape is small, ${\cal C}(S_{\rm ghost}, S_{\rm ortho}) = 0.25$, making it difficult to distinguish the bispectrum of ghost inflation from the equilateral bispectrum of the `pure $c_s$--theory'. 

\newpage
\section{Comments on Naturalness}
\label{sec:naturalness}

In this paper we have considered {\it weak coupling} as a new criterium to narrow down the in principle vast space of effective theories of inflation~\cite{Cheung, Senatore:2010wk}.
So far, we have not required our theories to be natural.
This is in the same spirit as the analogous situation in the Standard Model, where an unnaturally light Higgs particle is introduced to keep the theory of massive gauge bosons weakly coupled.
Physics beyond the Standard Model is required to explain the small Higgs mass. Similarly, additional structures in the high-energy theory may be required to make our theories natural.  We implicitly assumed that this doesn't change the low-energy phenomenology at $\omega < \omega_{\rm new}$.
In the effective theory of inflation \cite{Cheung, Senatore:2010wk,  Senatore:2010jy, LeoWMAP5} {\it naturalness} was proposed as a basic criterium to focus on the interesting regimes of the parameter space of couplings.
In this section, we comment briefly on how this notion of naturalness may be modified in our weakly-coupled UV-completions.

\vskip 4pt

Let us first review the argument of \cite{LeoWMAP5} concerning the natural values of parameters in the effective theory with small $c_s$.  Our starting point is the action (\ref{unitarygauge}), with $M_2^{4} \approx \tfrac{1}{2}M_{\rm pl}^2 |\dot H| c_s^{-2}$ and a priori unknown values for the coefficients $M_n^4$ for $n > 2$.  The interaction $M_3^4 (\delta g^{00})^3 = M_3^4 (\dot \pi^3 + \cdots)$ will be generated from $M_2^4 (\delta g^{00})^2 = M_2^4 (\dot \pi^2 + \dot \pi \frac{(\partial_i \pi)^2}{a^2} + \cdots)$ via the following loop 
\beq
\label{eqn:M3loop}
\includegraphicsbox[scale=.65]{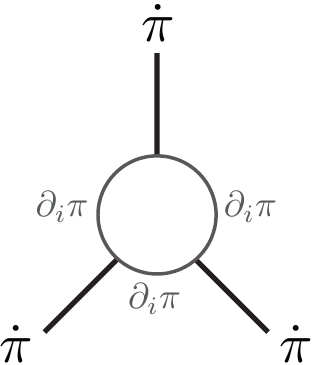}  \ \ =\ \ \dot \pi^3 \int \d \omega\, \d^3 k \, \frac{k^6}{(\omega^2 - c_s^2 k^2)^3} \ =\ \frac{\Lambda_{\rm uv}^4}{c_s^9} \, \dot \pi^3 \ \equiv\ M_3^4  \, \dot \pi^3\ ,
\eeq
where $\Lambda_{\rm uv}$ is the UV-cutoff of the loop integral.
If there is {\it no} new physics before the strong coupling scale $\Lambda_\star$, the UV-cutoff is at least $\Lambda_\star$.
In this case, loop effects generate a large value for $M_3$,
\beq
\label{equ:largeM3}
\Lambda_{\rm uv}^4 \to \Lambda_\star^4= M_2^4 c_s^7 \quad \Rightarrow \quad M_3^4 \sim \frac{M_2^4}{c_s^2}\ .
\eeq
As we have seen above, such a large value for $M_3$ can be of observational relevance since it is a prerequisite for generating an orthogonal shape for the bispectrum: $M_3^4 \approx - 5.4\, M_2^4\, c_s^{-2}$~\cite{LeoWMAP5}.

This renormalization argument is modified in our weakly-coupled examples.
Both of our UV-completions are characterized by a change in the dispersion relation from $\omega = c_s k$ to $\omega = k^2 / \rho$ at $\omega_{\rm new} = c_s^2 \rho$.  
This changes the high-energy behavior of the loop generating $M_3$.
With the new dispersion relation, the loop integral in (\ref{eqn:M3loop}) scales as
\beq
\includegraphicsbox[scale=.65]{feynman} \ \ = \ \ \dot \pi^3 \int_\Lambda \d\omega\, \d^3 k\, \frac{k^6}{\big(\omega^2 - \frac{k^4}{ \rho^2} \big)^3} \ \sim\  \frac{ \rho^{9/2}}{\Lambda^{1/2}}\, \dot \pi^3 \ =\ \frac{1}{c_s^9} \frac{\omega_{\rm new}^{9/2}}{\Lambda^{1/2}}\, \dot \pi^3 \ .
\eeq
The integral is now IR-dominated and hence converges in the UV.  The effective UV-cutoff of the low-energy theory becomes $\Lambda_{\rm uv}^4 \to \omega_{\rm new}^4$. In weakly-couped UV-completions we expect the scale of new physics to be parametrically below the scale of strong coupling.
In both of our examples we have seen this expectation confirmed: In the $\pi$-$\sigma$--model we found $\omega_{\rm new}^4 < c_s^3 \Lambda_\star^4$,
while the model with extrinsic curvature requires $\omega_{\rm new}^4 < c_s \Lambda_\star^4$.
This suppresses the loop-generated size of $M_3$,
\beq
M_3^4 \sim \frac{\omega_{\rm new}^4}{c_s^9} < \frac{M_2^4}{c_s^2}\ .
\eeq
A change in the dispersion relation below the strong coupling scale therefore naively seems to be a simple way to realize both weak coupling and a naturally small value of $M_3$.
However, what happens above the scale $\omega_{\rm new}$ in our specific examples is more complicated that just a change in the dispersion relation.  The physics responsible for changing the dispersion relation may also change the interaction Hamiltonian or even the scaling properties of the theory.  Therefore, we must consider the full UV-completion to determine the natural values of the low-energy parameters within a given model.

In any case, it should be clear that in our UV-completions there is nothing special about the scale $\Lambda_\star$.  As a result, any divergent integrals will be cut off at $\Lambda_{\rm b}$.  While some divergences are regulated by a change in the dispersion relation, both models contain additional operators that were fine-tuned.  This is most clear in the $\pi$-$\sigma$ model where we introduced a scalar $\sigma$ with a mass term $\mu^2 \sigma^2$.  Correction to $\mu^2$ are divergent and suggest that $\mu \sim \Lambda_{\rm b}$ in a natural theory.  Furthermore, one easily generates additional operators like $\tilde \mu \sigma^3$.  In this model, the question of fine-tuning has been moved to a scalar potential, but has not been resolved.  One would suspect that supersymmetry would resolve the problem, but constructing supersymmetric completions of our theories lies beyond the scope of this work.

\section{Discussion}
\label{sec:Conclusions}

With the increasing precision of CMB measurements by Planck and other experiments on the horizon, it is timely to ask what we can hope to learn about inflation from these observations.  In the absence of measurable B-mode polarization or deviations from Gaussianity, the amount of information is limited to the amplitude and the scale-dependence of the power spectrum of primordial density fluctuations.  In contrast, if non-Gaussianity were detected, it would provide an entire function worth of information about the physics of inflation~\cite{Komatsu:2009kd}.  

A powerful way to describe non-Gaussianity in single-field models is the effective theory of inflation~\cite{Cheung}.
In this approach, large interactions are associated with a small sound speed. The dominant signals arise from two distinct operators both of which produce equilateral shapes of non-Gaussianity. 
By fine-tuning the relative coefficients of these two operators one creates the so-called orthogonal shape~\cite{LeoWMAP5}.
In this paper, we discussed the physical implications of a detection of non-Gaussianity with approximately equilateral shape.
We showed that the associated effective theories become strongly coupled far below the symmetry breaking scale, but not far above the Hubble scale.
We compared this situation to the Standard Model of particle physics, where WW scattering becomes strongly coupled around the TeV scale, unless `new physics', such as a light Higgs particle, is introduced below the would-be strong coupling scale.
Similarly, 
for the effective theory of inflation to be weakly coupled at all energies, we require new physics to appear below the strong coupling scale.
Since the scale of new physics can't be too far above the Hubble scale, one may hope that it doesn't completely decouple from measurements of primordial non-Gaussianity.
We computed the signatures of candidate UV-completions perturbatively in the small ratio $H^2/\omega_{\rm new}^2$.
Interestingly, we found that the leading corrections to the bispectrum have a significant contribution in the squashed limit, and can therefore, in principle, be distinguished from the purely equilateral signal.
In practice, detecting the new physics requires a high-significance detection of the dominant equilateral signal and assumes that $H^2/\omega_{\rm new}^2$ is not too small.
Since observational constraints put a rather strong lower bound on $H^2/\omega_{\rm new}^2$, even the minimal correction to the equilateral shape may not be totally out of reach of future experiments.

\vskip 4pt
A number of future studies suggest themselves:
\begin{itemize}
\item[-] Although our analytic computations were performed for $H^2/\omega_{\rm new}^2 \ll 1$, it is more natural for $H^2/\omega_{\rm new}^2$ to be of order one.
Computing the shape of the three-point function in this case would be particularly interesting, but would require a non-perturbative treatment. 
Using our perturbative calculations as a guide, we may suspect that the bispectrum for $\omega_{\rm new}  \sim H$ could deviate significantly from the equilateral shape.  Checking this intuition explicitly will likely require numerical work. We plan to return to this question in the future.

\item[-] By demanding that inflationary models with small sound speed are weakly coupled at all energies, we were lead to consider fairly novel effective theories.  However, there was nothing special about considering small sound speed. In fact, new physics near the Hubble scale may appear in many other models with large non-Gaussianities---e.g.~the multi-field effective theory \cite{Senatore:2010wk} contains a large number of additional interactions, many of which lead to strong coupling near the Hubble scale.  
It would be interesting to extend our analysis to these cases and investigate the observational signatures of their UV-completions.

\item[-] Our conclusions aren't restricted to the bispectrum.  For example, a large trispectrum generated by the operator $M_4^4 (g^{00}+1)^4$~\cite{Senatore:2010jy} also implies a strong coupling scale close to the Hubble scale.  Again, if new physics becomes important near the Hubble scale, it may not decouple from observations.  The nature of the UV-completions that give rise to these large values of $M_4$ may differ from those relevant to small $c_s$, warranting a separate analysis.

\item[-] We have not required our theories to be technically natural. Constructing technically natural versions of our UV-completions could lead to additional structures at high energies. 
Again, this is similar to the situation in the Standard Model where an unnaturally light Higgs is made natural by supersymmetry.    It is interesting to note that supersymmetric models of inflation generically include particles with Hubble scale masses. Hence, supersymmetry could potentially offer a compelling explanation of `new physics' at the Hubble scale.  It would be worthwhile to develop supersymmetric versions of our UV-completions to either confirm that the additional physics decouples or to explore its low-energy signatures. 
\end{itemize}

\subsubsection*{Acknowledgements}

We are grateful to Nathaniel Craig, Thomas Dumitrescu, Raphael Flauger, Zohar Komargodski, Enrico Pajer, Rafael Porto, Soo-Jong Rey, Leonardo Senatore, Amit Yadav and Matias Zaldarriaga for helpful discussions.
The research of D.B.~is supported by the National Science Foundation under PHY-0855425 and a William D.~Loughlin Fellowship at the Institute for Advanced Study.
D.B.~thanks the Institute for Theoretical Physics, Madrid and the theory group at Cornell for hospitality.
The research of D.G.~is supported by the DOE under grant number DE-FG02-90ER40542 and the Martin A.~and Helen Chooljian Membership at the Institute for Advanced Study.  D.G.~thanks the Stanford Institute for Theoretical Physics for hospitality.
D.B.~and D.G.~thank the Centro de Ciencias de Benasque Pedro Pascual for hospitality
while this work was being completed. 

\newpage
\appendix

\section{Strong Coupling in DBI Inflation}
\label{sec:DBI}

In this paper, we studied the effective theory of {\it fluctuations} around quasi-de Sitter backgrounds.
We showed that in the limit of small sound speed these fluctuations become strongly coupled not far from the energy scale associated with the cosmological experiment, the Hubble scale.
In this appendix, we relate these findings to the well-known special properties of the effective theory of the {\it background}~\cite{Creminelli:2003iq,Weinberg:2008hq}.

\vskip 4pt
Inducing a non-trivial sound speed for the fluctuations requires higher-derivative terms to be dynamically important during inflation,
\beq
\label{equ:deri}
{\cal L} = - \frac{1}{2}(\partial_\mu \phi)^2 \left( 1 + \sum_{n=1}^\infty \frac{(\partial_\mu \phi)^{2n}}{M^{4n}} \right)\ .
\eeq
In fact, the limit $c_s \ll 1$ requires $(\partial_\mu \phi)^2 \sim M^4$~\cite{Creminelli:2003iq}, indicating a breakdown of the standard derivative expansion of the effective theory for the background. 
To make sense of the derivative expansion in (\ref{equ:deri}) seems to require a UV-completion.
A famous example for the UV-completion of models with small $c_s$ is Dirac-Born-Infeld (DBI) inflation~\cite{DBI}, 
\beq
\label{equ:DBIaction}
\mathcal{L}  = - M^4 \sqrt{1 - \frac{(\partial_{\mu} \phi)^2}{M^4}} \ .
\eeq
Although all terms in the expansion in powers of $(\partial_\mu \phi)^2$ are equally important in the limit $c_s \ll 1$, the higher-dimensional boost symmetry of the DBI action nevertheless controls the theory \cite{Tseytlin:1999dj}.
DBI inflation therefore is an example where the effective theory of the small $c_s$--background is well-defined.

What about the corresponding effective theory for the fluctuations?
What determines the strong coupling scale that we identified in the main text?  Because the strong coupling scale is expected to be above the Hubble scale, we can work in the Minkowski limit. 
Consider the DBI action (\ref{equ:DBIaction}).
Any background of the form $\dot{\bar \phi} = const.$~is a solution to the equations of motion.  We expand the action in fluctuations around the background, $\phi = \bar{\phi}(t) + \varphi(t, {\bf x})$, and choose $\dot{\bar{ \phi}} = M^2 \sqrt{1- c_s^2}$.  The action for the fluctuations is 
\beq
\mathcal{L}  = - M^4 c_s \sqrt{1 - 2(1-c_s^2)^{1/2} \frac{\dot\varphi}{c_s^2 M^2} - \frac{(\partial_{\mu} \varphi)^2}{c_s^2 M^4}} \ .
\eeq
Expanding the square root, we find the quadratic action
\beq
\mathcal{L}_2 = \frac{1}{2 c_s^3} (\dot \varphi^2- c_s^2 (\partial_i \varphi)^2) \ .
\eeq
For sufficiently large $\dot{\bar{\phi}}$, a small sound speed, $c_s \ll 1$, is generated for the fluctuations.  However, the expansion of the square root was only valid when $ \dot \varphi^2 < M^4 c_s^4 (1-c_s^2)^{-1}$.  Moreover, we were justified in treating $\varphi$ as the fluctuation around a background as long as $\dot\varphi^2 < M^4$.  Hence, the theory for the fluctuations is strongly coupled when
\beq
M^4 c_s^4 (1-c_s^2)^{-1} \ < \ \dot \varphi^2 \ <\ M^4 \ .
\eeq
 This range of energies is consistent with our previous analysis.
 In particular, as before, the strong coupling scale is suppressed by four powers of $c_s$ relative to the symmetry breaking scale.
 
\newpage
\section{Dynamics of the $\pi$-$\sigma$ Model}
\label{sec:dynamics}

In this appendix, we describe the dynamics of the $\pi$-$\sigma$ model. Our starting point is the quadratic action 
\beq
{\cal L}_2 = - \frac{1}{2}(\partial_\mu \pi_c)^2 - \frac{1}{2}(\partial_\mu \sigma)^2 + \rho\, \dot \pi_c \sigma - \frac{1}{2}\mu^2 \sigma^2\ .
\eeq
The corresponding equations of motion are
\bea
\ddot \pi + 3 H \dot \pi + k_{\rm p}^2 \,\pi &=& -\rho \bigl[ 3H \sigma + \dot \sigma \bigr] \ , \\
\ddot \sigma + 3 H \dot \sigma + k_{\rm p}^2 \, \sigma &=&  -\mu^2 \sigma + \rho\, \dot \pi\ ,
\eea
where $k_{\rm p} = k/a(t)$ is the physical momentum.  Since we will always be working in the limit $\rho \gg H$, terms proportional to $\rho$ become important well inside the horizon.  The dynamics in the regime $k \sim \rho$ are therefore well-approximated by the flat space limit\footnote{Including the effects of finite $H$ using a WKB-like approximation is straightforward, but doesn't qualitatively affect our arguments.  } $H \to 0$.  In this limit, we can find the mode solutions exactly using the ansatz $\pi = A e^{i \omega t}$ and $\sigma = B e^{i \omega t}$.  The equations of motion become algebraic equations for $A$, $B$ and $\omega$,
\bea
\left[ \omega^2- k^2\, \right] \pi &=&  \rho \left( i \omega \right) \sigma \ , \\
\left[ \omega^2 - (k^2+ \mu^2)\, \right] \sigma &=& - i \rho \omega \, \pi  \ .
\eea
Combining these two equations, we find 
\beq
\label{equ:B6}
\omega^2_\pm \ =\ k^2 + \frac{ \rho^2 + \mu^2}{2} \pm \sqrt{\frac{(\rho^2 +\mu^2)^2}{4} + \rho^2 k^2} \ .
\eeq
When $k \gg \rho > \mu$, we have two positive frequency modes with $\omega \sim k$.  This is not surprising.  At very high energies $\pi$ and $\sigma$ are essentially independent free fields.  
When $k \ll \rho$, we expand the square root in (\ref{equ:B6}) to find
\beq
\omega_{\pm}^2\ \simeq \ k^2 + \frac{ \rho^2 + \mu^2}{2} \pm \left( \frac{\rho^2 +\mu^2}{2} + \frac{\rho^2}{\rho^2 +\mu^2} \, k^2 - \frac{\rho^4}{(\rho^2+\mu^2)^3} \, k^4+\cdots \right)\ .
\eeq
The $\omega_{+}$ solution gives rise to a positive frequency mode with $\omega \sim \rho$.  This describes a very massive degree of freedom as its energy is always of order $\rho$.  
In contrast, the positive $\omega_{-}$ solution corresponds to a positive frequency mode with $\omega \ll \rho$, i.e.~a light degree of freedom.  The dispersion relation for this mode is 
\beq
\label{equ:dis}
\omega_{-}^2 \ \simeq\ \left(1- \frac{\rho^2}{\rho^2 + \mu^2} \right)  k^2 + \frac{\rho^4}{(\mu^2+\rho^2)^3} \, k^4 \ \equiv\ c_s^2 k^2 + \frac{k^4}{\tilde{\rho}^2}\ ,
\eeq
where $c_s^2 = \mu^2 / (\rho^2+ \mu^2)$ and $\tilde \rho^2 = (\rho^2 + \mu^2)^3 / \rho^4$.

We are interested in the behavior for $\mu^2 \ll \rho^2$.  In this case, $c_s^2 \approx \mu^2/\rho^2 \ll 1$ and $\tilde \rho \approx \rho$.  When $ \rho > k > c_s \rho$, we see that the second term in (\ref{equ:dis}) dominates, and  the dispersion relation is $\omega \approx k^2 / \rho$.  Therefore, for the range of energies $ \rho > \omega > c_s^2 \rho$ the mode is effectively described by the free Schr\"odinger equation.  When $\omega < c_s^2 \rho \equiv \omega_{\rm new}$ (or $k< \mu$), we return to a linear dispersion relation with a small speed of sound, $\omega \approx c_s k$.   

As we have seen, when $\omega < \rho$, there is only one degree of freedom.  In this limit, the full quadratic action is unnecessary as it includes the heavier mode.  The action that describes just the light mode is 
\beq
{\cal L}_2  \approx \rho\, \dot \pi_c \sigma - \frac{1}{2}\frac{(\partial_i \pi_c)^2}{a^2} - \frac{1}{2}\frac{(\partial_i \sigma)^2}{a^2} - \frac{1}{2}\mu^2 \sigma^2\ .
\eeq
Here we have dropped the relativistic kinetic terms, $\dot \pi^2_c$ and $\dot \sigma^2$, as they introduce corrections which at low energies are suppressed by $\omega / \rho$.  The equations of motion now are,
\bea
 k_{\rm p}^2 \,\pi &=&- \rho \bigl[ 3H \sigma + \dot \sigma \bigr] \ , \label{equ:pi1} \\
 k_{\rm p}^2\, \sigma &=&  - \mu^2 \sigma + \rho\, \dot \pi\ . \label{equ:sig1}
\eea
Repeating our analysis in the $H \to 0$ limit, we find that
\beq
\omega^2 = \frac{\mu^2}{\rho^2} \, k^2 + \frac{k^4}{\rho^2} \ .
\eeq
When $\mu \ll \rho$, this reproduces the dispersion relation of the full quadratic action (\ref{equ:dis}).

\newpage
\section{Corrections to Vanilla Sound Speed Models}
\label{sec:Observables}

In this appendix, we describe in more detail the predictions of our two-field UV-completion of small speed of sound theories.
We will consider the limit $\omega_{\rm new} \equiv \mu^2/\rho > H$, where the corrections to the `pure $c_s$--theory', ${\cal L}_0 \equiv {\cal L}_{\rm sr} + \tfrac{1}{2} M_2^4 (\delta g^{00})^2$, can be treated perturbatively.

\subsection{Preliminaries}

Our starting point will be the low-energy effective action for the Goldstone mode in the decoupling limit
\beq
\label{equ:EffectiveAction}
{\cal L}_{\rm eff} = M_{\rm pl}^2 |\dot H| \left[  - (\partial_\mu \pi)^2 + \frac{\rho^2}{\mu^2} \left( \dot \pi - \tfrac{1}{2} (\partial_\mu \pi)^2 \right) \left( 1 + \frac{\Box}{\mu^2} + \cdots \right) \left( \dot \pi - \tfrac{1}{2} (\partial_\mu \pi)^2 \right) \right] + \cdots\ ,
\eeq
where
\beq
c_s^{-2} \equiv 1 + \frac{\rho^2}{\mu^2}\ .
\eeq
We will compute correlation functions of the comoving curvature perturbation
$\zeta = - H \pi$.
The quadratic Lagrangian for $\zeta$ is
\bea
\label{equ:free}
L_2 \equiv a^4 {\cal L}_2 &=&  {\cal C} \cdot \frac{(a H)^2}{c_s^2} \Bigl[ (\zeta')^2 - c_s^2 (\partial_i \zeta)^2 \Bigr]\ ,
\eea
where ${\cal C} \equiv  \frac{M_{\rm pl}^2 |\dot H|}{H^4}$.
For quasi-De Sitter backgrounds, ${\cal C} \approx const.$ and $(aH) \approx - \tau^{-1}$.
The leading interactions in the limit $c_s \ll 1$ are
\bea
L_{\rm int}^{(0)} &=&  {\cal C}\cdot \frac{aH}{c_s^2} \cdot\zeta' (\partial_i \zeta)^2\ , \\
L_{\rm int}^{(1)} &=&  {\cal C} \cdot \frac{H^2}{c_s^2 \mu^2}  \cdot \zeta' \partial_i^2 \zeta'\ , \\
L_{\rm int}^{(2)} &=&  {\cal C}  \cdot \frac{H^2}{c_s^2 \mu^2} \cdot \frac{aH}{c_s^2} \cdot\zeta' \,\frac{c_s^2 \partial_j^2}{(aH)^2}\, (\partial_i \zeta)^2\ .
\eea
The interaction Hamiltonian is $H_{\rm int} = - \int \d^3 x\, L_{\rm int}$.
We promote the field $\zeta$ to the operator $\hat \zeta$, whose Fourier modes we expand in creation and annihilation operators
\beq
 \hat \zeta_{\bf k}(\tau) = \zeta_k(\tau) \hat a_{\bf k} + \zeta_k^*(\tau) \hat a^\dagger_{- {\bf k}}\ ,
\eeq
where
\beq
[\hat a_{{\bf k}}, \hat a^\dagger_{-{\bf k}'}] = (2\pi)^3\, \delta({\bf k} + {\bf k}')\ .
\eeq
We implicitly treat $\hat \zeta$ as interaction picture fields whose time-evolution is determined by $H_0 = - \int \d^3 x\, L_2$. 
We will use the in-in formalism to compute correlation functions (for a recent review see \cite{Chen:2010xk})
\beq
\label{equ:inin}
\langle \hat Q \rangle(\tau) = \langle 0 | \left[ \bar{\rm T} \, e^{i \int_{-\infty}^\tau \d \tau' \hat H_{\rm int}(\tau')}\right] \hat Q(\tau)  \left[ {\rm T} \, e^{- i \int_{-\infty}^\tau \d \tau' \hat H_{\rm int}(\tau')}\right] | 0 \rangle \ ,
 \eeq
 where $|0 \rangle$ is the vacuum of the free theory, $\hat a_{\bf k} |0\rangle \equiv 0$, and the symbols ${\rm T}$ and $\bar {\rm T}$ denote time-ordering and anti-time-ordering, respectively.
 To compute equal time $n$-point functions of $\zeta$, we let
  $\hat Q = \{\, \hat \zeta_{{\bf k}_1} \hat \zeta_{{\bf k}_2} \, , \, \hat \zeta_{{\bf k}_1} \hat \zeta_{{\bf k}_2} \hat \zeta_{{\bf k}_3}\, , \, \cdots \}$.
Expanding the exponentials in (\ref{equ:inin}) in powers of $\hat H_{\rm int}$ allows us to compute correlation functions  perturbatively.  
We evaluate each term in the series using contractions and normal ordering.
After normal ordering, the only terms that are non-vanishing are those with all terms contracted. A contraction between two terms, $\hat \zeta_{{\bf k}}(\tau')$ (on the left) and $\hat \zeta_{\bf q}(\tau'')$ (on the right), gives
\beq
\acontraction[0.5ex]{:}{\hat{\zeta}_{\bf k}}{(\tau')}{\hat{\zeta}_{\bf q}}
:\hat \zeta_{\bf k}(\tau')\hat \zeta_{\bf q}(\tau''): \ \ =\  [\hat \zeta_{\bf k}(\tau'), \hat \zeta_{\bf q}(\tau'') ] = \zeta_k(\tau') \zeta_q^*(\tau'')\, \delta({\bf k} + {\bf q})\ .
\eeq
Feynman diagrams are a convenient way of keeping track of all necessary contractions.
The final integrals will be highly oscillatory in the infinite past due to form of the mode functions $\propto e^{-i c_s k \tau}$.
To evaluate the integral we therefore perform a Wick rotation $\tau \to i \tau$.

\subsection{Two-Point Function}

The free-field action (\ref{equ:free}) implies the standard mode functions in de Sitter space
\beq
\label{equ:modefct}
\zeta_k(\tau) = \zeta_k^{(o)} \cdot e^{-ic_s k \tau} (1 + i c_s k \tau)\ ,
\eeq
where
\beq
 k^{3/2} \zeta_k^{(o)} \equiv \frac{H}{M_{\rm pl}} \frac{i}{\sqrt{ 4 c_s \epsilon}} \ .
\eeq
This allows us to compute the two-point function (or power spectrum) after horizon crossing, $P_\zeta(k) =  |\zeta_k^{(o)}|^2$, where
\beq
\langle \hat \zeta_{{\bf k}_1} \hat \zeta_{{\bf k}_2} \rangle = (2\pi)^3\, P_\zeta(k_1) \, \delta({{\bf k}_1} + {\bf k}_2)\ .
\eeq
The dimensionless power spectrum  is
\beq
\Delta_\zeta \equiv k^{3} P_\zeta = \frac{1}{4} \frac{H^2}{M_{\rm pl}^2 c_s \epsilon} = \frac{1}{4} \frac{1}{{\cal C}\, c_s}\ .
\eeq
From the quadratic correction term $H_{\rm int}^{(1)}$ we get a tree-level correction to the power spectrum
\beq
\lim_{\tau \to 0}\ \langle \hat \zeta_{{\bf k}_1}\hat \zeta_{{\bf k}_2} \rangle (\tau) = - i \int_{-\infty}^0 \d \tau' \ \langle [  \hat \zeta_{{\bf k}_1} \hat \zeta_{{\bf k}_2}(0) , \hat H_{\rm int}^{(1)}(\tau')]\rangle\ ,
\eeq
or
\beq
\Delta P_\zeta(k) =-i \cdot {\cal C} \cdot \frac{H^2}{c_s^2 \mu^2} \times\zeta_{k}^{(o)} \zeta_{k}^{(o)} \times 2\int_{-\infty}^0 \d \tau'  \left[ k^2 \frac{d \zeta^*_{k}}{d\tau'}  \frac{d \zeta^*_{k}}{d\tau'} \right] \ +\ c.c. 
\eeq
Using the mode functions (\ref{equ:modefct}) and performing a Wick rotation to regulate the integral, we find 
\beq
\Delta P_{\zeta} = \frac{1}{4} \frac{H^2}{\mu^2 c_s^2} \times P_{\zeta}\ .
\eeq
We deduce that $H_{\rm int}^{(1)}$ simply induces an unobservable shift of the amplitude of the power spectrum.

\subsection{Three-Point Function}

Next, we compute the three-point function (or bispectrum)
\beq
\langle \hat \zeta_{{\bf k}_1}  \hat \zeta_{{\bf k}_2}  \hat \zeta_{{\bf k}_3} \rangle \equiv (2\pi)^3\, B_\zeta(k_1,k_2,k_3)\,  \delta({\bf k}_1 + {\bf k}_2+{\bf k}_3) \ .
\eeq
The leading bispectrum, corresponding to the `pure $c_s$--theory', is
\beq
\lim_{\tau \to 0}\ \langle \hat \zeta_{{\bf k}_1} \hat \zeta_{{\bf k}_2} \hat \zeta_{{\bf k}_3} \rangle (\tau)  = - i \int_{-\infty}^0 \d \tau' \langle [  \hat \zeta_{{\bf k}_1} \hat \zeta_{{\bf k}_2}   \hat \zeta_{{\bf k}_3}(0) , \hat H_{\rm int}^{(0)}(\tau')]\rangle\ .
\eeq
Substituting $\hat H_{\rm int}^{(0)}$, we find 
\bea
\lim_{\tau \to 0}\ \langle \hat \zeta_{{\bf k}_1} \hat \zeta_{{\bf k}_2} \hat \zeta_{{\bf k}_3} \rangle (\tau) &=& i \cdot {\cal C} \cdot \frac{1}{c_s^2} \cdot (2\pi)^3 \, \delta({\bf k}_1 + {\bf k}_2 + {\bf k}_3) \times \zeta_{k_1}^{(o)}  \zeta_{k_2}^{(o)} \zeta_{k_3}^{(o)} \times \\ 
&& \tfrac{1}{2}(k_1^2 - k_2^2 - k_3^2) \times \int \frac{\d \tau'}{\tau'} \, (\zeta_{k_1}^*)' \zeta_{k_2}^* \zeta_{k_3}^*\ +\ perms. \ + \ c.c. \nonumber
\eea
This gives~\cite{LeoWMAP5} 
\beq
B^{(0)}_\zeta = \frac{1}{4} \frac{\Delta_\zeta^2}{c_s^2} \cdot \frac{K_1^{\, 6}  - 3 K_1^{\, 4}  K_2^{\, 2}  + 11 K_1^{\, 3} K_3^{\, 3}   - 4 K_1^{\, 2}  K_2^{\, 4}    - 4 K_1 K_2^{\, 2} K_3^{\, 3}   +  12 K_3^{\, 6}    }{K_3^{\, 9} K_1^{\, 3}}\ ,
\eeq
where 
\bea
K_1 &=& k_1 +k_2+k_3 \ ,\\
K_2 &=& (k_1 k_2 +k_2 k_3 + k_3 k_1)^{1/2} \ ,\\
K_3 &=& (k_1 k_2 k_3)^{1/3}\ .
\eea

To compute the correction induced by a combination of  $H_{\rm int}^{(0)}$  and $H_{\rm int}^{(1)}$, we find it convenient to use an alternative form of (\ref{equ:inin}),
\bea
\lim_{\tau \to 0}\ \langle \hat \zeta_{{\bf k}_1}  \hat \zeta_{{\bf k}_2}  \hat \zeta_{{\bf k}_3}  \rangle (\tau) &=& \int_{-\infty}^0 \d \tau' \int_{-\infty}^{\tau'} \d \tau'' \langle 0 | \hat H_{\rm int}(\tau')  \hat \zeta_{{\bf k}_1}  \hat \zeta_{{\bf k}_2}  \hat \zeta_{{\bf k}_3}(0) \hat H_{\rm int}(\tau'') | 0\rangle  \nonumber \\
&-&  \int_{-\infty}^0 \d \tau' \int_{-\infty}^{\tau'} \d \tau'' \langle 0 | \hat H_{\rm int}(\tau'') \hat H_{\rm int}(\tau')  \hat \zeta_{{\bf k}_1}  \hat \zeta_{{\bf k}_2}  \hat \zeta_{{\bf k}_3}(0) | 0\rangle  \nonumber \\
&-&  \int_{-\infty}^0 \d \tau' \int_{-\infty}^{\tau'} \d \tau'' \langle 0 |  \hat \zeta_{{\bf k}_1}  \hat \zeta_{{\bf k}_2}  \hat \zeta_{{\bf k}_3}(0) \hat H_{\rm int}(\tau') \hat H_{\rm int}(\tau'')| 0\rangle  + \cdots \ , \label{equ:3pt}
\eea
where $\hat H_{\rm int} \equiv \hat H_{\rm int}^{(0)}  + \hat H_{\rm int}^{(1)}$.
The leading corrections come from terms with one factor of $\hat H_{\rm int}^{(1)}$  and one factor of $\hat H_{\rm int}^{(2)}$. 
The first integral in (\ref{equ:3pt}) therefore is  
\bea
 \int_{-\infty}^0 \d \tau' \int_{-\infty}^0 \d \tau'' \ \langle 0 | \hat H_{\rm int}^{(0)}(\tau') \hat \zeta_{{\bf k}_1}  \hat \zeta_{{\bf k}_2}  \hat \zeta_{{\bf k}_3}(0) \hat H_{\rm int}^{(1)}(\tau'') | 0\rangle \ +\ c.c. &=& \\
 &&\hspace{-10cm} =\ \ \ \  \, \frac{{\cal C}^2}{c_s^2} \cdot \frac{H^2}{c_s^2 \mu^2} \cdot (2\pi)^3 \, \delta({\bf k}_1 + {\bf k}_2 + {\bf k}_3)  \times  \zeta_{k_1}^{(o)} \zeta_{k_2}^{(o)} \zeta_{k_3}^{(o)}  \times  \nonumber  \\
&& \hspace{-10cm } \tfrac{1}{2}(k_3^2 - k_1^2 -k_2^2) k_3^2 \times \int_{-\infty}^0 \frac{\d \tau'}{\tau'} \zeta_{k_1} \zeta_{k_2} \zeta_{k_3}'  \int_{-\infty}^0 \d \tau'' (\zeta'_{k_3})^* (\zeta'_{k_3})^* \ +\ perms. \ + \ c.c. \nonumber
\eea
The second integral in (\ref{equ:3pt}) is 
\bea
 && - \int_{-\infty}^0 \d \tau' \int_{-\infty}^{\tau'} \d \tau'' \ \langle 0 | \left[ \hat H_{\rm int}^{(0)}(\tau'') \hat H_{\rm int}^{(1)}(\tau') +  \hat H_{\rm int}^{(1)}(\tau'') \hat H_{\rm int}^{(0)}(\tau') \right] \hat \zeta_{{\bf k}_1}  \hat \zeta_{{\bf k}_2}  \hat \zeta_{{\bf k}_3}(0)  | 0\rangle  \ =  \\
 && \hspace{-0.8cm} = \hspace{1.5cm} \, \frac{{\cal C}^2}{c_s^2} \cdot \frac{H^2}{c_s^2 \mu^2} \cdot (2\pi)^3 \, \delta({\bf k}_1 + {\bf k}_2 + {\bf k}_3) \times \zeta_{k_1}^{(o)} \zeta_{k_2}^{(o)} \zeta_{k_3}^{(o)}  \times   \tfrac{1}{2}(k_3^2 - k_1^2 -k_2^2) k_3^2\  \times  \nonumber \\
 && \hspace{-0.5cm} \left[ \int_{-\infty}^0 \d \tau' (\zeta'_{k_3})^* \zeta'_{k_3}   \int_{-\infty}^{\tau'} \frac{\d \tau''}{\tau''}  \zeta_{k_1} \zeta_{k_2} \zeta_{k_3}' \ \, + \right.  \left.  \int_{-\infty}^0 \frac{\d \tau'}{\tau'}   \zeta_{k_1} \zeta_{k_2} (\zeta_{k_3}')^*   \int_{-\infty}^{\tau'} \d \tau''  \zeta'_{k_3} \zeta'_{k_3} \ \, + \ perms. \ \right]\, . \nonumber
\eea
Finally, the third integral in (\ref{equ:3pt}) is the complex conjugate of the second integral.
The sum of the three integrals has a rather complex analytic answer, $B_\zeta^{(1)}$. Showing this answer here wouldn't be very illuminating.

Finally, there is a tree-level correction to the bispectrum from the interaction $H_{\rm int}^{(2)}$,
\beq
\lim_{\tau \to 0}\ \langle \hat \zeta_{{\bf k}_1} \hat \zeta_{{\bf k}_2} \hat \zeta_{{\bf k}_3} \rangle (\tau) = - i \int_{-\infty}^0 \d \tau' \ \langle [  \hat \zeta_{{\bf k}_1} \hat \zeta_{{\bf k}_2} \hat \zeta_{{\bf k}_3}(0) , \hat H_{\rm int}^{(2)}(\tau')]\rangle\ .
\eeq
Substituting $\hat H_{\rm int}^{(2)}$, we find
\bea
\lim_{\tau \to 0}\ \langle \hat \zeta_{{\bf k}_1} \hat \zeta_{{\bf k}_2} \hat \zeta_{{\bf k}_3} \rangle (\tau) &=& i \cdot {\cal C} \cdot \frac{H^2}{c_s^2 \mu^2} \cdot (2\pi)^3 \, \delta({\bf k}_1 + {\bf k}_2 + {\bf k}_3) \times \zeta_{k_1}^{(o)}  \zeta_{k_2}^{(o)} \zeta_{k_3}^{(o)} \times \\ 
&& \tfrac{1}{2}(k_1^2 - k_2^2 - k_3^2)\, k_1^2 \times \int_{-\infty}^0 \d \tau' \, \tau' \, (\zeta_{k_1}^*)' \zeta_{k_2}^* \zeta_{k_3}^*\ +\ perms. \ + \ c.c. \nonumber
\eea
Performing the same manipulations as before, we get 
\bea
&& B_\zeta^{(2)} \ =\  \frac{1}{18}\, \frac{H^2}{c_s^2 \mu^2} \frac{\Delta_\zeta^2}{c_s^2} \cdot \frac{K_1^{\, 8}  - 3 K_1^{\, 6} K_2^{\, 2}  - 7 K_1^{\, 4} K_2^{\, 4} +12 K_1^{\, 2} K_2^{\, 6} + 17 K_1^{\, 5} K_3^{\, 3}   }{K_3^{\, 9} K_1^{\, 5}} \, \times \nonumber \\
&& \hspace{3.2cm}\times \
 \frac{ - 43 K_1^{\, 3} K_2^{\, 2} K_3^{\, 3} + 36 K_1 K_2^{\, 4} K_3^{\, 3} + 66 K_1^{\, 2} K_{3}^{\, 6} - 48 K_2^{\, 2} K_3^{\, 6}      }{K_3^{\, 9} K_1^{\, 5}}\ .
\eea
In \S\ref{sec:explicit} we discuss the shapes of all bispectra computed in this appendix.

\newpage
\section{Small $\mu$ Limit of the $\pi$-$\sigma$ Model}
\label{sec:muZero}

In this appendix, we compute the power spectrum and the bispectrum in the $\mu \to 0$ limit of the $\pi$-$\sigma$ model.
As we showed in Appendix~\ref{sec:dynamics}, the dynamics of the Goldstone mode in this limit are characterized by a non-linear dispersion relation $\omega \sim k^2/\rho$. The theory is therefore similar, but, as we will show, not identical to ghost inflation~\cite{GhostInflation}.
We also argued in Appendix~\ref{sec:dynamics} that the theory at the energies relevant for inflation, $\omega \sim H \ll \rho$, is described by a single degree of freedom. However, the single-field action in this limit is non-local, so we will find it more convenient to consider the local two-field action
\beq
{\cal L} \ \approx \ \rho\, \dot \pi_c \sigma - \frac{1}{2} \frac{(\partial_i \pi_c)^2}{a^2} - \frac{1}{2}\frac{(\partial_i \sigma)^2}{a^2}  \underbrace{- \frac{1}{\xi} (\partial_\mu \pi_c)^2 \sigma}_{{\cal L}_{\rm int}}\ ,
\eeq
where $\xi \equiv \frac{2 M_{\rm pl}^2 |\dot H|}{m^3} = \frac{2}{\rho}\, (2 M_{\rm pl}^2 |\dot H|)^{1/2}$.
The quadratic part of the action, ${\cal L}_2$, implies the following equations of motion
\beq
\label{equ:EOM}
 \ddot \pi_c + 5 H \dot \pi_c + \frac{k^4}{\rho^2 a^4}\, \pi_c = 0 \qquad \,{\rm and} \, \qquad \frac{k^2}{a^2} \, \sigma =  \rho\, \dot{\pi}_c\ .
\eeq
In this appendix, we quantize the theory and compute the power spectrum and bispectrum of curvature fluctuations $\zeta = - H\pi$.

\subsection{Canonical Quantization}

At low energies, $\omega \ll \rho$, the coupling between $\pi_c$ and $\sigma$ dominates the dynamics. The field $\sigma$ then plays the role of the conjugate momentum of $\pi_c$, 
\beq
p_\pi \equiv \frac{\partial {\cal L}'}{\partial \dot \pi_c} =  \rho \sigma \ .
\eeq
 The equal-time canonical commutation relation
 \beq
\bigl[ \hat \pi_c({\bf x}, \tau) , \hat p_\pi({\bf y}, \tau) \bigr] = i a^{-3}\, \delta({\bf x} - {\bf y})\ ,
 \eeq
 then implies
 \beq
 \label{commutator}
 \bigl[ \hat \pi_c({\bf k}, \tau) , \hat \sigma({\bf q}, \tau) \bigr] = \frac{i  }{a^3 \rho}\,  (2\pi)^3\, \delta({\bf k} + {\bf q})\ .
 \eeq
This confirms that the fields $\pi_c$ and $\sigma$ are not independent degrees of freedom at low energies.
 In particular, at low energies, $\sigma$ is proportional to $\pi_c'$,
 \beq
 \label{equ:SIG}
 \frac{k^2}{a^2} \, \sigma =  \frac{\rho}{a} \,\pi_c'\ .
 \eeq
 To obtain an analytic solution to the equation of motion (\ref{equ:EOM}), we use conformal time and define $v \equiv a^2 \pi_c$, such that
 \beq
 v'' + \left( \kappa^4 \tau^2 - \frac{6}{\tau^2} \right) v = 0 \ , \qquad {\rm where} \quad \kappa^2 \equiv \frac{k^2 H}{\rho}\ .
 \eeq
 This has a solution in terms of Hankel functions,
 \beq
 \label{solution}
 v =  (-\tau)^{1/2} \left[ c_1(\kappa)\, H_{5/4}^{(1)} \Bigl( \tfrac{1}{2}(\kappa\tau)^2 \Bigr) + c_2(k)\, H_{5/4}^{(2)} \Bigl( \tfrac{1}{2}(\kappa\tau)^2 \Bigr)\right]\ .
 \eeq
Our unusual commutation relations imply that some care is required to define the correct Bunch-Davies vacuum.
From (\ref{commutator}) and (\ref{equ:SIG}) we obtain
 \beq
 \label{comm2}
 [ \hat v({\bf k}, \tau) , \hat v'({\bf q}, \tau)] = i \, \frac{k^2}{\rho^2}\, (2\pi)^3\, \delta({\bf k}+ {\bf q})\ .
 \eeq
 This implies a non-standard normalization for the mode functions
 \beq
 \label{NORM}
 v_k v^{*'}_{k} - v^{*}_{k} v_{k}' = i\, \frac{k^2}{\rho^2}\ ,
 \eeq
 if we define the standard operator mode expansion
 \beq
 \hat v({\bf k}, \tau) = v_k(\tau) \hat a_{\bf k} + v^*_k(\tau) \hat a^\dagger_{- {\bf k}}\ ,
 \eeq
with  $[ a_{\bf k} , a^\dagger_{- {\bf q}}] = (2\pi)^3 \delta({\bf k} + {\bf q})$.
Substituting (\ref{solution}) into (\ref{NORM}) gives 
 \beq
 \label{norm}
 \frac{8}{\pi} \Big[ |c_1(k)|^2 -|c_2(k)|^2\Big] = \frac{k^2}{\rho^2}\ .
 \eeq
Selecting the positive frequency solutions at early times fixes $c_2 = 0$, and hence
 \beq
c_1(k) = \sqrt{\frac{\pi}{8}} \, \frac{k}{\rho}\ .
 \eeq
 The mode functions are then completely determined 
  \beq
 v_k(\tau) = \sqrt{\frac{\pi}{8}} \frac{k}{\rho}  (-\tau)^{1/2}\, H_{5/4}^{(1)}\Bigl(\tfrac{1}{2}\tfrac{H}{\rho} (k \tau)^2 \Bigr)\ .
 \eeq
In the superhorizon limit this becomes
 \beq
v^{(o)}_k \equiv \lim_{k \tau \to 0} v_k \, \sim\, a^2 k^{-3/2} (\rho H^3)^{1/4}\ .
 \eeq

\subsection{Two-Point Function}

 Using $v^2 = a^4 \pi_c^2 = 2 \hskip 1pt a^4 M_{\rm pl}^2 \epsilon\, \zeta^2$, we arrive at the power spectrum for $\zeta$ after horizon crossing
 \beq
 \Delta_\zeta = k^3 |\zeta_k^{(o)}|^2 \sim \frac{H^2}{M_{\rm pl}^2 \epsilon}  \left( \frac{\rho}{H}\right)^{1/2}  \ .
 \eeq
 We note that the spectrum is scale-invariant.
 The amplitude is enhanced by a factor of $\rho/H \gg 1$ relative to the familiar expression from slow-roll inflation.
 
\subsection{Three-Point Function}

The three-point function in the small $\mu$ limit is quite interesting.  The interactions involve new operators that contain both $\pi$ and $\sigma$ fields.  Because $\sigma$ is the canonical momentum, it is not clear what to expect from these interactions.

A rough measure of the expected size of non-Gaussianity is
 \beq
 \frac{{\cal L}_3}{{\cal L}_2} \sim \frac{\xi^{-1} (\partial_\mu \pi_c)^2 \sigma }{\rho\, \dot \pi_c \sigma} \sim \frac{1}{\xi\, \rho}\frac{(\partial_i \pi_c)^2}{\dot \pi_c} \sim \frac{k^2}{\omega} \frac{\pi_c}{(M_{\rm pl}^2 |\dot H|)^{1/2}} \sim \rho \, \pi \sim \frac{\rho}{H}\, \zeta\ ,
 \eeq
 or
 \beq
 f_{\rm NL} \sim \frac{1}{\zeta} \frac{{\cal L}_3}{{\cal L}_2} \sim \frac{\rho}{H} \gg 1 \ .
 \eeq
 In the regime of interest this is a detectably large amount of non-Gaussianity.
We compute the detailed shape of the bispectrum using the formalism of Appendix~\ref{sec:Observables},
\beq
\lim_{\tau \to 0}\ \langle \hat \zeta_{{\bf k}_1} \hat \zeta_{{\bf k}_2} \hat \zeta_{{\bf k}_3} \rangle = - i \int_{-\infty}^0 \d \tau' \langle [ \hat \zeta_{{\bf k}_1} \hat \zeta_{{\bf k}_2} \hat \zeta_{{\bf k}_3}(0) , \hat H_{\rm int}(\tau')] \rangle\ ,
\eeq
where
\beq
H_{\rm int} = - \int \d^3 x\, a^4\, {\cal L}_{\rm int} \approx - \frac{2 {\cal C}}{\xi} \int \d^3 x\, (a H)^2 (\partial_i \zeta)^2 \sigma\ .
\eeq
The bispectrum is then determined by the following integral
\bea
\lim_{\tau \to 0}\ \langle \hat \zeta_{{\bf k}_1} \hat \zeta_{{\bf k}_2} \hat \zeta_{{\bf k}_3} \rangle (\tau) &=& i \cdot {\cal C} \cdot \frac{\rho^2}{H^2} \cdot (2\pi)^3 \, \delta({\bf k}_1 + {\bf k}_2 + {\bf k}_3) \times \zeta_{k_1}^{(o)}  \zeta_{k_2}^{(o)} \zeta_{k_3}^{(o)} \times \\ 
&&\frac{ \tfrac{1}{2}(k_1^2 - k_2^2 - k_3^2)}{k_1^2} \times \int_{-\infty}^0 \frac{\d \tau'}{(\tau')^3}  \, (\zeta_{k_1}^*)' \zeta_{k_2}^* \zeta_{k_3}^*\ +\ perms. \ + \ c.c. \nonumber
\eea
where we used
\beq
\frac{\sigma_k}{\xi} = \frac{1}{2} \frac{aH}{k^2} \cdot \frac{\rho^2}{H^2} \cdot \zeta_k'\ .
\eeq
We find that the shape generated by this interaction is almost indistinguishable from the equilateral shape generated by $M_2^4 \dot \pi (\partial_i \pi)^2$ in the small $c_s$--theory (see \S\ref{sec:muZeroX}).

\newpage
\section{Unitarity Bounds}
\label{sec:unitarity}

In this appendix, we derive unitarity bounds for the models discussed in the paper.

\subsection{Preliminaries}

Unitarity of the S-matrix translates into a statement about amplitudes through the optical theorem~\cite{Peskin}:
\bea
\label{eqn:unitarity}
2\, {\rm Im}\, \mathcal{A}(k_1, k_2 \to p_1, p_2) &=& \nonumber \\
&&  \hspace{-4.5cm} =\ \prod_i \int \frac{\d^{3} q_i}{(2\pi)^3} \frac{1}{2 \omega_i}  \mathcal{A}(k_1,k_2 \to \{ q_i \})  \mathcal{A}^{*}(p_1,p_2 \to \{ q_i \})\, \delta^{(4)}\Big(k_1 + k_2 - \sum_i q_i \Big)\ .
\eea
This expression will be most useful for  $2 \to 2$ scattering, in which case the matrix elements on both sides of the optical theorem are the same.  It is convenient to rewrite the amplitude for $2 \to 2$ scattering in the center of mass frame and expand it in Legrendre polynomials
\beq
\label{equ:Adef}
\mathcal{A}(k_1, k_2 \to p_1, p_2)_{\rm cm} = \left(\frac{\partial k}{\partial \omega} \frac{k^2}{\omega^2} \right)^{-1}16 \pi \sum_\ell (2 \ell +1) P_\ell (\cos\theta)\, a_\ell\ ,
\eeq
where $\cos \theta = \frac{1}{2}\hat {\bf p}_1 \cdot \hat {\bf p}_2$.  The subscript $\ell$ labels the spin-$\ell$ partial wave. The prefactor in brackets is unity in a relativistic theory, but will play an important role in our non-relativistic examples.  Due to angular momentum conservation, we can write (\ref{eqn:unitarity}) as
\beq
\label{eqn:partialwave}
{\rm Im} \hskip 2pt a_{\ell} =| \, a_{\ell} \, |^2\ .
\eeq
This expression leads to a powerful constraint, because it cannot be satisfied if $({\rm Re} \hskip 2pt a_{\ell})^2 > \frac{1}{4}$. A violation of (\ref{eqn:partialwave}) implies a violation of unitarity.  In practice, the leading contributions to ${\rm Re} \hskip 2pt a_{\ell}$ are computed perturbatively.  When these contributions to ${\rm Re} \hskip 2pt a_{\ell}$ exceed $\frac{1}{2}$, one concludes that the theory is strongly coupled, i.e.~higher-order terms must be of equal importance for the result to be consistent with unitarity.  Let's see how this works in our examples.  

\subsection{Unitarity and Small Sound Speed}

We will start with the `pure $c_s$-theory', ${\cal L}_{\rm sr} + \frac{1}{2} M_2^4 (\delta g^{00})^2$.  We will consider the unitarity bound associated with the operator $\frac{1}{2} M_2^4 (\partial_i \pi \partial^i \pi)^2$ which, after canonical-normalization, becomes
\beq
\label{equ:INT}
{\cal L}_{\rm int} = \frac{1}{8} \frac{(1-c_s^2) c_s^2}{2 M_{\rm pl}^2 |\dot H|} (\partial_i \pi_c \partial^i \pi_c)^2 \ .
\eeq
For convenience, we have extracted a factor of $\frac{1}{8}$ since it equals the combinatorial factor arising from this interaction.  The strength of the interaction is controlled by the dimensionless ratio $\omega^4 / (2\hskip 1pt M_{\rm pl}^2 \dot H c_s^5)$.  We expect the theory to become strongly coupled at some order-one value of this ratio.  We will use the perturbative violation of (\ref{eqn:partialwave}) to compute this number. 

The amplitude generated by the interaction (\ref{equ:INT}) is
\beq
\mathcal{A}(k_1, k_2 \to p_1, p_2)_{\rm cm} = \frac{(1-c_s^2) }{2 M_{\rm pl}^2 |\dot H| c_s^2 } \,\omega^4 \left[ 1 + 2 \cos ^2(\theta) \right] \ .
\eeq
We compare this to (\ref{equ:Adef}), which now reads
\beq
\mathcal{A}(k_1, k_2 \to p_1, p_2)_{\rm cm} = c_s^3 \cdot 16 \pi\sum_\ell (2 \ell +1) P_\ell (\cos\theta)\, a_\ell \ .
\eeq
The largest partial wave amplitude is the s-wave component
\beq
a_0 = \frac{1}{16 \pi} \frac{(1-c_s^2) \omega^4 }{2 M_{\rm pl}^2 |\dot H| c_s^5 } \int \d\cos\theta\, \left[1+ 2 \cos^2 \theta\right] = \frac{1}{4 \pi} \frac{(1-c_s^2) \omega^4 }{2 M_{\rm pl}^2 |\dot H| c_s^5 } \ .
\eeq
Using $a_0 < \frac{1}{2}$ we find
\beq
\omega^4 <  4 \pi \hskip 1pt M_{\rm pl}^2 |\dot H| \hskip 1pt \frac{c_s^5}{1-c_s^2}\ .
\eeq

\subsection{The Extrinsic Curvature Model}

Let us apply the same reasoning to our UV-completion with extrinsic curvature terms (cf.~\S\ref{sec:Extrinsic}).
As in the small $c_s$--theory, the strong coupling scale is determined by the operator $\frac{1}{2} M_2^4 (\partial_i \pi \partial^i \pi)^2$.  After canonical-normalization, the interaction is 
\beq
\label{equ:INT2}
{\cal L}_{\rm int} = \frac{1}{8} \frac{1}{4 M_{2}^4} (\partial_i \pi_c \partial^i \pi_c)^2 \ .
\eeq
Due to the modified dispersion relation, the strength of the interaction is controlled by the dimensionless ratio $\omega^{1/2} / (M_2^4 \rho^{7/2})$.  We will compute the order-one value of this ratio at the energy scale at which the theory becomes strongly coupled.

The amplitude for $2 \to 2$ scattering generated by the interaction (\ref{equ:INT2}) is 
\beq
\mathcal{A}(k_1, k_2 \to p_1, p_2)_{\rm cm} = \frac{\rho^2}{4 M_2^4 }\, \omega^2 \left[ 1 + 2 \cos ^2\theta \right]\ .
\eeq
This answer is essentially the same as in the small $c_s$--case, except that we have used a different dispersion relation to write the amplitude in terms of $\omega$ alone.  Again, the largest partial wave amplitude is the s-wave component
\beq
a_0 = \frac{1}{16 \pi} \frac{ \rho^ {7/2} \omega^{1 / 2} }{ M_{2}^4} \ .
\eeq 
Using (\ref{eqn:partialwave}), we require $a_0 < 1/2$ for the S-matrix to be unitary. This implies
\beq
\label{eqn:unitarityE}
\omega < (8 \pi)^2 \frac{M_2^8}{\rho^7} \ .
\eeq
The numerical factor in (\ref{eqn:unitarityE}) is larger than in the `pure $c_s$--theory' because the operator has a  smaller scaling dimension.

\subsection{The $\pi$-$\sigma$ Model}

Finally, we wish to determine the scale at which the $\pi$-$\sigma$ model (\S\ref{sec:pisig}) becomes strongly coupled.  Unlike the previous two examples, perturbative unitarity is not a useful measure of strong coupling.  As we will show, infrared divergences associated with the massless Goldstone bosons make the analysis ill-defined.  We will instead define strong coupling by estimating the size of loop corrections directly.  

The largest coupling in the $\pi$-$\sigma$ model arises from the interaction $m^3 \partial_i \pi \partial^i \pi \hskip 2pt \sigma$.  After canonical-normalization, the interaction becomes
\beq
\label{equ:INT3}
{\cal L}_{\rm int} = \frac{1}{2}\frac{1}{(2 M_{\rm pl} |\dot H|)^{1/2}\rho^{1/2}} \, \partial_i \pi_c \partial^i \pi_c\hskip 2pt \sigma_c \ .
\eeq
We are interested in $2 \to 2$ scattering of the Goldstone bosons.
Because $\sigma$ is not an independent field, we need to be careful about properly defining the propagator.  Completing the square as usual, we find that the propagator takes the form
\bea
\langle \Psi^i \Psi_j \rangle = \frac{1}{\omega^2 - k^4/\rho^2} \left( \begin{array}{cc}
\frac{k^2}{\rho} & - i \omega  \\
i \omega & \frac{k^2}{\rho} 
\end{array} \right)\ ,
\eea
where $\Psi^{i} \equiv ( \sigma \, \, \pi)$.  We note that on shell $\sigma = \rho \dot \pi/k^2$.  Since the external lines are on shell, the Feynman rules for the external lines involving $\sigma$ are identical to those of $\pi$.
Using Feynman rules, we can therefore compute the $2\to2$ scattering amplitude.  The full amplitude is not important, as we only need to understand the $\theta \to 0$ limit 
\beq
\mathcal{A}(k_1, k_2 \to p_1 = k_1, p_2 = k_2)_{\rm cm} \simeq \frac{2}{\theta^2} \frac{ \omega^3 \rho}{M_{\rm pl}^2 |\dot H|} \ .
\eeq
The divergence as $\theta \to 0$ is associated with the t-channel exchange of a massless particle with vanishing energy and momentum.  This divergence is common in a theory with massless particles and reflects the fact that the probability for final states with a finite number of particles is zero.
Unfortunately, the divergence also undermines our ability to use the partial wave expansion to determine the strong coupling scale.  These infrared divergences infect our observables and prevent us from isolating the UV-behavior.  

We will therefore define strong coupling as the scale where one-loop and tree-level contributions to the same process are equal.  Because (\ref{equ:INT3}) is a three-particle interaction, for every loop integral we need two insertions of the coupling, and hence get a suppression by $16  \pi^2$.  Finally, we get the following bound
\beq
\omega^{3/2} < 16 \pi^2 \, \frac{2 M_{\rm pl}^2 |\dot H|}{\rho^{5/2}} \ .
\eeq
The additional powers of $\rho$ come from the momentum integrals in every loop.  This is equivalent to defining a dimensionless coupling by using the dispersion relation.

\newpage
 \begingroup\raggedright\endgroup

\end{document}